\pdfoutput=1
\documentclass[11pt]{article}

\usepackage[]{acl}

\usepackage{times}
\usepackage{latexsym}

\usepackage[T1]{fontenc}
\usepackage{array}
\usepackage[utf8]{inputenc}

\usepackage{microtype}

\usepackage{inconsolata}
\usepackage{CJKutf8}
\usepackage{graphicx}

\usepackage{color, xcolor}

\usepackage[utf8]{inputenc}
\usepackage{booktabs}
\usepackage{algorithm}
\usepackage{algorithmic}
\usepackage{multicol}
\usepackage{multirow, makecell, caption}
\usepackage{colortbl}
\usepackage{tikz}
\usepackage{arydshln}
\usepackage{pgfplots}
\usepackage{amsmath}
\usepackage{amssymb}
\usepackage{tcolorbox}
\usepackage{tabularx}
\usepackage{enumerate}
\usepackage{pifont}

\makeatletter
\def\adl@drawiv#1#2#3{%
        \hskip.5\tabcolsep
        \xleaders#3{#2.5\@tempdimb #1{1}#2.5\@tempdimb}%
                #2\z@ plus1fil minus1fil\relax
        \hskip.5\tabcolsep}
\newcommand{\cdashlinelr}[1]{%
  \noalign{\vskip\aboverulesep
           \global\let\@dashdrawstore\adl@draw
           \global\let\adl@draw\adl@drawiv}
  \cdashline{#1}
  \noalign{\global\let\adl@draw\@dashdrawstore
           \vskip\belowrulesep}}
\makeatother

\title{
ChemHTS: Hierarchical Tool Stacking for Enhancing Chemical Agents}

\author{
\textbf{Zhucong Li\textsuperscript{1}\thanks{\enspace Equal contribution.}},
 \textbf{Jin Xiao\textsuperscript{3}\footnotemark[1]},
 \textbf{Bowei Zhang\textsuperscript{3}},
  \textbf{Zhijian Zhou\textsuperscript{1}},
 \\
 \textbf{Qianyu He\textsuperscript{2}},
\textbf{Fenglei Cao\textsuperscript{4}},
 \textbf{Jiaqing Liang\textsuperscript{3}},
  \textbf{Yuan Qi\textsuperscript{1}\thanks{\enspace Corresponding author.}}
  \\
  \textsuperscript{1}Artificial Intelligence Innovation and Incubation Institute, Fudan University
 \\
 \textsuperscript{2}School of Computer Science, Fudan University
 \\
 \textsuperscript{3}School of Data Science, Fudan University
 \\
 \textsuperscript{4}Shanghai Academy of Artificial Intelligence for Science
 \\
    \{zcli22, jinxiao23, bwzhang24\}@m.fudan.edu.cn, \\
    \{liangjiaqing, qiyuan\}@fudan.edu.cn
}

\begin{document}
\thispagestyle{empty}
\maketitle

\begin{abstract}
Large Language Models (LLMs) have demonstrated remarkable potential in scientific research, particularly in chemistry-related tasks such as molecular design, reaction prediction, and property estimation. While tool-augmented LLMs have been introduced to enhance reasoning and computation in these domains, existing approaches suffer from tool invocation errors and lack effective collaboration among diverse tools, limiting their overall performance. To address these challenges, we propose ChemHTS (Chemical Hierarchical Tool Stacking), a novel method that optimizes tool invocation pathways through a hierarchical stacking strategy. ChemHTS consists of two key stages: tool self-stacking warmup and multi-layer decision optimization, enabling LLMs to refine tool usage dynamically. We evaluate ChemHTS across four classical chemistry tasks and demonstrate its superiority over strong baselines, including GPT-4o, DeepSeek-R1, and chemistry-specific models, including ChemDFM. Furthermore, we define four distinct tool-stacking behaviors to enhance interpretability, providing insights into the effectiveness of tool collaboration.
Our dataset and code are publicly available at \url{https://github.com/Chang-pw/ChemHTS}.

\end{abstract}

\section{Introduction}


\begin{figure}[t] 
    \centering
        \includegraphics[width=0.45\textwidth]{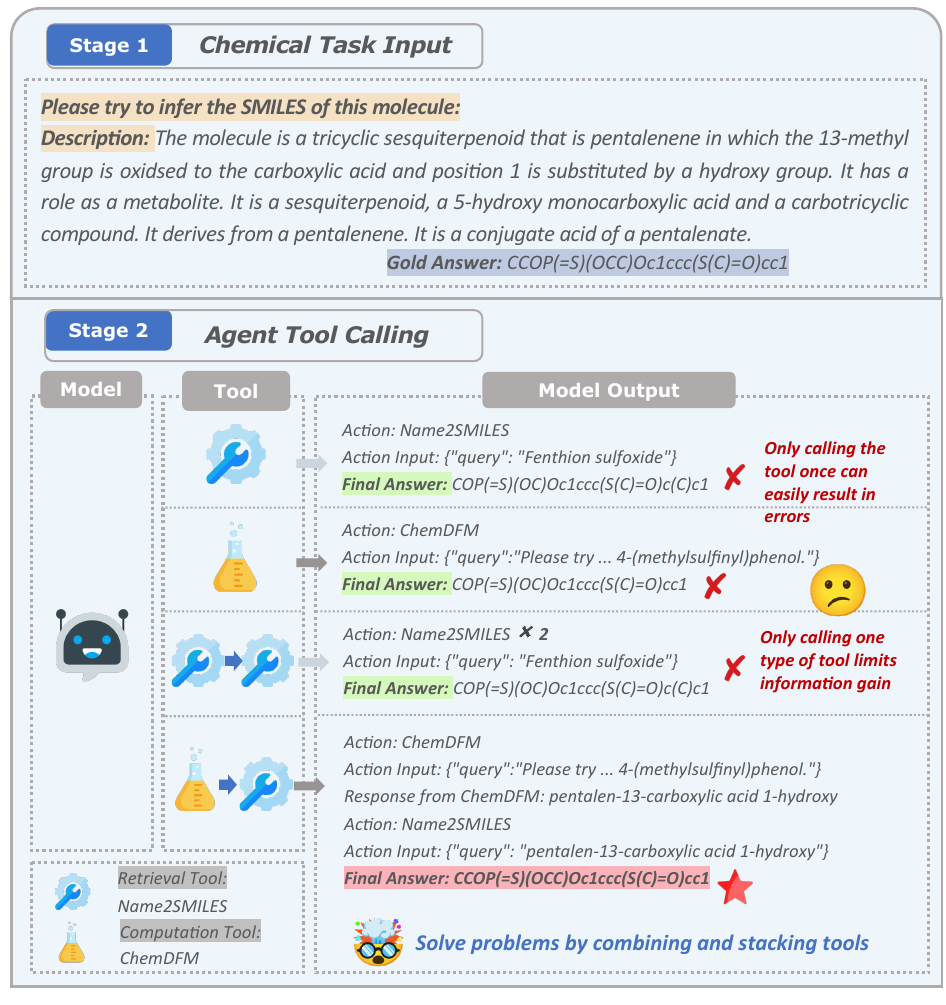}
    \caption{
    Using the text-based molecule design task as an example, this analysis examines the issues in model tool usage under Name2SMILES and ChemDFM tools.}
    \label{fig:intro}
\end{figure}

\begin{figure*}[t] 
    \centering
    \includegraphics[width=0.95\textwidth]{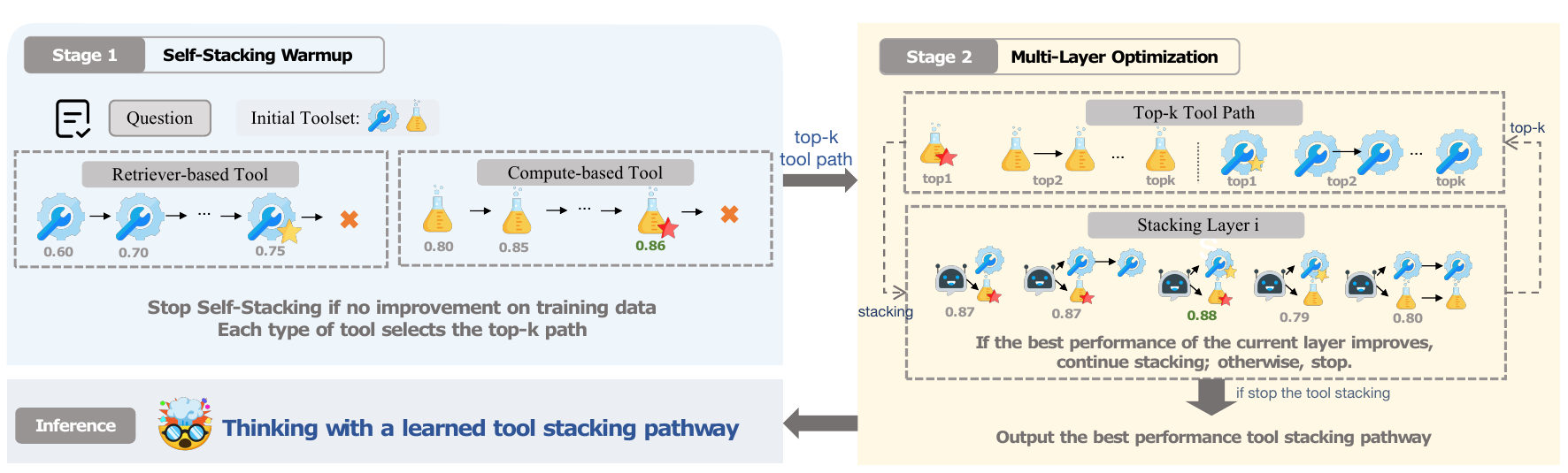}
    \caption{
    Our ChemHTS method framework diagram. For each chemical task, we identify the optimal tool-stacking pathway through the ChemHTS method for subsequent task inference execution.
    }
    \label{fig:010method}
\end{figure*}

In recent years, large language models (LLMs) ~\cite{touvron2023llama,achiam2023gpt,deepseekai2025deepseekr1incentivizingreasoningcapability} have made groundbreaking advancements in fields such as natural language processing (NLP) and knowledge reasoning, and are gradually being applied to scientific research, particularly demonstrating immense potential in the domain of chemistry~\cite{guo2023largelanguagemodelschemistry,ouyang2024structuredchemistryreasoninglarge}. LLMs leverage their robust capabilities for knowledge integration and reasoning computation to offer novel solutions for tasks such as molecular design~\cite{noutahi2023gottasafenewframework}, reaction prediction~\cite{shi2023relmleveraginglanguagemodels}, and property prediction~\cite{srinivas2024crossmodallearningchemistryproperty}. This progress significantly enhances the level of automation and intelligence in chemical research.

Despite the promising prospects of LLMs in the field of chemistry, current research faces challenges. Existing models exhibit limitations in their ability to generalize across diverse tasks, and they lack access to external knowledge sources, which constrains their practical utility in scientific applications~\cite{zhao2024chemdfmlargelanguagefoundation,liao2024wordsmoleculessurveylarge,han2024generalistspecialistsurveylarge}. Consequently, studies have integrated expert-designed tools to enhance the performance of LLMs in chemical tasks~\cite{bran2023chemcrowaugmentinglargelanguagemodels,boiko2023autonomous,song2024multi}.


However, chemical tool-augmented LLMs still exhibit numerous challenges when confronted with complex and diverse toolsets~\cite{bran2023chemcrowaugmentinglargelanguagemodels}.
Firstly, due to the limited capabilities of tools or the susceptibility of LLMs to tool calling errors, relying on a single tool invocation may result in biased outcomes~\cite{ye2024toolswordunveilingsafetyissues}. As illustrated in Fig.~\ref{fig:intro}, if the result of invoking an agent tool once is biased, the model is unable to self-correct and may produce incorrect answers.
Secondly, LLMs struggle to collaborate across different types of tools, limiting their capacity to maximize information gain~\cite{cheng2025toolunlearningtoolaugmentedllms}. During reasoning, they often depend on a single category of tools, overlooking the synergistic benefits of integrating multiple tool types. As a result, errors can propagate through the reasoning chain, ultimately impairing overall task performance.

Despite the emergence of numerous studies in recent years focused on optimizing tool usage, these efforts have not yet adequately addressed the aforementioned issues. Current research primarily concentrates on optimal tool selection for individual tasks, while overlooking the collaborative interactions between tools and their impact on task inference capabilities.
Some studies have attempted to leverage the context learning capabilities of LLMs by employing strategic prompting to optimize tool selection. For instance, frameworks such as Chain of Thought (CoT) ~\cite{wei2023chainofthoughtpromptingelicitsreasoning} and ReACT ~\cite{yao2023reactsynergizingreasoningacting,yang2023autogptonlinedecisionmaking} aim to guide model in effectively utilizing tools during the reasoning process. 
Another category of methods involves fine-tuning LLMs on tool calling datasets to enhance their tool usage capabilities. For instance, approaches such as ToolBench~\cite{qin2023toolllmfacilitatinglargelanguage} utilizes supervised fine-tuning~\cite{shen2024rethinkingdataselectionsupervised} to improve the model's understanding and calling of tools. 
However, these approaches can only enhance the selection of individual tools to a limited extent and have not succeeded in achieving collaborative optimization among multiple tools.

To address the aforementioned issues, we propose a novel tool calling optimization method termed \textbf{ChemHTS} (\textbf{Chem}ical \textbf{H}ierarchical \textbf{T}ool \textbf{S}tacking), 
which aims to enhance the tool invocation process of LLMs in chemistry-related tasks. The core idea of C he m is to explore the optimal tool invocation pathways through a hierarchical stacking strategy under the constraints of a given chemical task and multiple tools, thereby improving the reasoning and computational capabilities of LLMs in chemical applications. This method encompasses two key stages: tool self-stacking and tool multi-layer decision optimization.

In addition, we apply the ChemHTS method to four classic chemistry tasks: text-based molecular design, molecular description, molecular property prediction, and reaction prediction. We systematically explore the optimal tool invocation pathways for each task and conduct extensive experiments on the test sets.
The results demonstrate that the tool invocation paths optimized by ChemHTS outperform several baseline models, such as GPT4o~\cite{openai2024gpt4ocard} and DeepSeek-R1~\cite{deepseekai2025deepseekr1incentivizingreasoningcapability}, as well as chemistry-specific models, including UniMol~\cite{ji2024unimol2exploringmolecularpretraining} and ChemDFM~\cite{zhao2024chemdfmlargelanguagefoundation}, across all tasks. Furthermore, we define four distinct tool stacking behavioral patterns—\textbf{Correct, Modify, Judge, and Reserve}-and provide case studies to elucidate the reasons behind the improvements in task performance due to tool stacking.
To summarize, our contributions are mainly three-fold: 

\begin{itemize}
    \item We are the first to highlight the advantages of agent tool collaboration and propose the ChemHTS method to optimize the optimal tool stacking path for agents across various chemical tasks, addressing the issue of low efficiency in the utilization of agent tools.
    \item Through extensive experiments, we have demonstrated that the optimal tool stacking paths derived from our method outperform numerous baseline models and chemistry-specific models across four classical chemical tasks, thereby validating the effectiveness and generalizability of our approach.
    \item We define four behavioural patterns of models during the tool stacking process, to conduct an in-depth analysis and interpretation of the reasons behind the performance improvements in tool stacking, thereby enhancing the interpretability of the experiments.
\end{itemize}

\section{Related Work}
\subsection{Large Language Models for Chemistry}

Large language models (LLMs) have demonstrated significant potential in chemistry, with applications spanning molecular generation, property prediction, reaction modeling, and retrosynthetic analysis~\cite{fang2024molinstructionslargescalebiomolecularinstruction,tang2024prioritizingsafeguardingautonomyrisks,liao2024wordsmoleculessurveylarge}. For instance, the ChemDFM ~\cite{zhao2024chemdfmlargelanguagefoundation} pretrained on chemical literature and textbooks and further refined through extensive instruction tuning, has exhibited enhanced performance across various chemical tasks. Similarly, ChemCrow~\cite{bran2023chemcrowaugmentinglargelanguagemodels}, an LLM-powered chemistry assistant, integrates multiple expert-designed chemical tools to improve LLM performance in chemistry-related applications.Despite these advancements, LLMs continue to face challenges in handling complex chemical computations and generalizing across diverse chemical problems~\cite{ouyang2024structuredchemistryreasoninglarge,han2024generalistspecialistsurveylarge}. Moreover, they remain inefficient in utilizing existing computational chemistry tools~\cite{shi2023relmleveraginglanguagemodels}, and struggle to navigate the combinatorial and hierarchical relationships between these tools.

\subsection{Tool-augmented LLMs}
LLMs~\cite{anil2023palm,achiam2023gpt,touvron2023llama} have demonstrated strong reasoning capabilities in natural language processing and scientific computing. However, they face limitations in specialized tasks in fields such as chemistry and physics~\cite{yang2024moosechemlargelanguagemodels}, including constrained computational accuracy, insufficient numerical reasoning abilities, and a lack of collaboration with external tools. To address these shortcomings, researchers have recently proposed the tool-augmented LLMs approach~\cite{qin2023toolllmfacilitatinglargelanguage,wang2024gtabenchmarkgeneraltool,yang2023autogptonlinedecisionmaking}, enabling LLMs to dynamically call external tools and thereby enhance their task execution capabilities.
Representative methods include ReAct~\cite{yao2023reactsynergizingreasoningacting}, which combines chain-of-thought reasoning (CoT)~\cite{wei2023chainofthoughtpromptingelicitsreasoning} with tool invocation to allow LLMs to dynamically acquire external information during decision-making, and Toolformer~\cite{schick2023toolformerlanguagemodelsteach}, which enables LLMs to autonomously decide when to call tools, improving the accuracy of computational tasks. 
Despite these advancements, existing research primarily focuses on single-tool invocation and has yet to explore hierarchical combinations of tools. A single tool is often insufficient to solve complex scientific problems, whereas the collaborative invocation of multiple tools holds promise for enhancing the reasoning capabilities of LLMs in chemical tasks.

\section{ChemHTS}
Our study proposes a hierarchical tool stacking method, named ChemHTS, to optimize the tool invocation process of LLMs in chemistry-related tasks. ChemHTS iteratively refines and optimizes the tool stacking trajectory, ultimately identifying the optimal sequence of tool calls to enhance computational performance (Fig.\ref{fig:010method}).
\subsection{Step 1: Tool Self-stacking Warmup}

To evaluate the performance of a single tool after multiple invocations and determine its suitability for stacking, we first conduct a self-stacking warm-up phase for the individual tool.
Since research in chemical tasks often integrates theoretical computations with experimental knowledge, leveraging computation to predict unknown properties or processes and retrieval to access existing knowledge, we categorize the chemical LLM toolset $\mathcal{L}$ into two types: compute-based tools $\mathcal{L}_c$ and retriever-based tools $\mathcal{L}_r$.
For computational tools, it may be necessary to test whether repeated invocations lead to improved computational accuracy or stability. In contrast, for retriever-based tools, it may be important to evaluate whether multiple retrievals can enhance knowledge coverage.

For a given chemical task $\mathcal{T}$ and its corresponding tool library $\mathcal{L}$, the performance of an individual tool ($ToolA_i$) and its self-stacking ($A_i \to A_i$) is evaluated. If repeated invocations of the same tool type result in performance degradation, further stacking of that tool is discontinued. From each category, the top-k most effective tool combinations are selected for the next stage of stacking, as outlined in Algorithm~\ref{alg:HTS}.. 
These refined tool-stacking pathways are encapsulated as new tools, reducing the subsequent search space, improving computational efficiency, and providing stronger tool combination strategies for the next phase.
\subsection{Step 2: Multi-Layer Optimization}
\begin{algorithm}[H]
\caption{ChemHTS Algorithm}
\small
\label{alg:HTS}
\begin{algorithmic}[1]
\REQUIRE Training dataset $\mathcal{D}$, tool library $\mathcal{L}$ (including compute-based tools $\mathcal{L}_c$ and retriever-based tools $\mathcal{L}_r$), maximum hierarchy depth $n$, number of selected paths per category $k$, maximum tool selection per step $m$.
\ENSURE Optimized tool invocation strategy $\mathcal{P}^*$.

\STATE \textbf{STEP 1: SELF-STACKING WARMUP}
\FOR{each tool $T_i \in \mathcal{L}_c \cup \mathcal{L}_r$}
    \STATE Evaluate self-stacking paths of $T_i$ on $\mathcal{D}$.
\ENDFOR
\STATE Select the Top-$k$ best-performing paths for computational tools $\mathcal{P}_c$ and retriever tools $\mathcal{P}_r$.
\STATE Encapsulate selected tool stacks as new tools in $\mathcal{L}$.

\STATE \textbf{STEP 2: MULTI-LAYER OPTIMAZATION}
\STATE Initialize hierarchy level $L \gets 1$.
\WHILE{$L \leq n$}
    \STATE Sort computational tools $\mathcal{L}_c$ and RAG tools $\mathcal{L}_r$ based on performance metrics.
    \STATE Select Top-1 computational tool $T_c^*$ and Top-$k$ RAG tools $\mathcal{P}_r$.
    \STATE Select Top-1 RAG tool $T_r^*$ and Top-$k$ computational tools $\mathcal{P}_c$.
    
    \FOR{each tool subset $\mathcal{L}_L \subseteq (\mathcal{P}_c \cup \mathcal{P}_r)$, $|\mathcal{L}_L| \leq m$}
        \STATE Evaluate the performance of the tool invocation path.
        \IF{current best performance $\leq$ previous best performance}
            \STATE \textbf{STOP} stacking process.
            \STATE \textbf{BREAK}.
        \ENDIF
    \ENDFOR
    \STATE Encapsulate best-performing paths as new tools.
    \STATE $L \gets L + 1$.
\ENDWHILE

\STATE Select the best tool invocation path $\mathcal{P}^*$ as the final strategy.
\RETURN $\mathcal{P}^*$.

\end{algorithmic}
\end{algorithm}



To explore the optimal invocation pathways across tool combinations and progressively encapsulate the best tools, ultimately determining the globally optimal tool invocation strategy, we divide the tool-stacking process into multiple layers and perform stacking optimization step by step.
In the first layer, the Top-1 computational tool selected in Stage 1 is combined with the Top-k retrieval tools, and the Top-1 retrieval tool is combined with the Top-k computational tools. Each agent is restricted to selecting the best tool from a pool of $m$ candidate tools in each step. If the optimal performance in this layer is worse than that of Stage 1, the stacking process is terminated; otherwise, the iteration continues.
In subsequent layers, all tools are re-ranked, and the top $k$ optimal pathways in each category are selected and encapsulated as new tools. This process continues until either the maximum number of layers $n$ is reached or no further optimization space remains.
Finally, the optimal tool invocation pathway is selected as the general tool invocation strategy for the given task.

\section{Experiments}

\subsection{Experiment Setup}

\paragraph{Dataset}


We evaluate the performance of ChemHTS in the field of chemistry using ChemLLMBench. ChemLLMBench~\cite{guo2023largelanguagemodelschemistry} comprises a series of chemistry-related tasks that cover a wide range of chemical topics. 
In this study, we focus on four representative tasks and select 100 evaluation instances for each task, consistent with the evaluation experiments in ChemDFM, as the test set.
Since the evaluation dataset for each task in ChemLLMBench contains only 100 instances, we adopt a similar approach to that in ~\cite{guo2023largelanguagemodelschemistry} to select the training set. 
For the \textbf{Text-Based Molecule Design and Molecule Captioning} tasks, we randomly sample 100 instances from the ChEBI-20-MM~\cite{liu2025quantitativeanalysisknowledgelearningpreferences}
dataset, excluding the corresponding test set, as the training set. 
For the \textbf{Reaction Prediction} task, we randomly sample 100 instances from the USPTO-MIT~\cite{jin2017predictingorganicreactionoutcomes} dataset, excluding the corresponding test set, as the training set. 
For the \textbf{Molecular Property Prediction} task, we randomly sample 50 instances for each dataset from the BBBP, HIV, BACE, Tox21, and ClinTox~\cite{wu2018moleculenetbenchmarkmolecularmachine} datasets, excluding the corresponding test sets, as the training set.
The details of our dataset are shown in Tab.~\ref{tab:dataset}.
\setlength\tabcolsep{1pt}
\begin{table*}[!htb]
    \centering
    \tiny
    \definecolor{lightgray}{gray}{0.9} 
    \resizebox{0.98\textwidth}{!}{
    \begin{tabular}{llllccll} 
    \hline
    \textbf{Ability} & \textbf{Task}                    & \textbf{Task Type} & \textbf{Dataset}                     & \textbf{\#train} & \textbf{\#test} & \textbf{Retrieval Tool} & \textbf{Compute Tool} \\ \hline
    Understanding    & Molecular Property Prediction    & Classification     & BBBP, HIV, BACE, Tox21, ClinTox      & 250              & 100             & SMILES2Property         & UniMol-v2 \\ \hline
    Reasoning        & Reaction Prediction              & Generation         & USPTO-MIT                            & 100              & 100             & SMILES2Property         & Chemformer \\ \hline
    Text-Based Molecule Design & Generation             & -                  & ChEBI-20-MM                          & 100              & 100             & Name2SMILES             & ChemDFM \\ \hline
    Explaining       & Molecule Captioning              & Generation         & ChEBI-20-MM                          & 100              & 100             & SMILES2Description      & Text+ChemT5 \\ \hline
    \end{tabular}
    }
    \caption{Details of the training and test sets for the four chemistry tasks.}
    \label{tab:dataset}
\end{table*}
\paragraph{Models}
We categorize current models into three primary groups: task-specific specialist models, LLM-based generalist models, and tool-based agent models. Task-specific specialist models refer to non-LLM models designed for specific tasks. Tool-based agent models leverage GPT-4o~\cite{openai2024gpt4ocard} as the core agent, augmented with additional tools tailored to the specific task. 
To ensure fair comparisons, we use the same test set for evaluating different models on each task.
\paragraph{Tools}
To facilitate the experimental process, for each task, we provide only the most relevant tools for the two categories: computational tools and retrieval tools. 
Details can be found in Tab.~\ref{tab:dataset}.

\subsection{Results}
\subsubsection{Text-based Molecule Design}
In the text-based molecule design task, LLMs predict a molecule’s SMILES (Simplified Molecular Input Line Entry System) representation based on a given description, testing their ability to interpret and translate chemical language into valid molecular structures~\cite{zhao2024chemdfmlargelanguagefoundation}.
Our study employs two sets of metrics to evaluate the performance of the task. The first set of metrics measures the text-based similarity between the predicted SMILES and the gold standard SMILES, including exact match, BLEU, and Levenshtein distance~\cite{haldar2011levenshteindistancetechniquedictionary}. 
The second set of metrics assesses the chemical similarity between the predicted molecules and the reference molecules, encompassing the validity of the predicted SMILES and the FTS (Fingerprint Tanimoto Similarity)~\cite{tanimoto1958elementary}, calculated based on MACCS, RDK, and Morgan~\cite{Morgan1965TheGO}.
\setlength\tabcolsep{1pt}
\begin{table}[!htb]
    \centering
    \tiny
    \definecolor{lightgray}{gray}{0.9} 
    \resizebox{0.5\textwidth}{!}{
    \begin{tabular}{lccccccc}
        \toprule
        \textbf{Model} & \textbf{Exact↑} & \textbf{BLEU↑} & \textbf{Dis↑} & \textbf{Validity↓} & \textbf{MACCS↑} & \textbf{RDK↑} & \textbf{Morgan↑} \\
        \midrule
        \rowcolor{lightgray} \multicolumn{8}{c}{\textit{Task-specific specialist models}} \\

        MolXPT~\cite{liu-etal-2023-molxpt} & 0.22 & - & - & \textbf{0.98} & 0.86 & 0.76 & 0.67 \\
        Text+Chem T5~\cite{textchemt5} & 0.32 & 0.85 & 16.87 & 0.94 & \underline{0.90} & \underline{0.82} & \underline{0.75} \\
        Mol-Instruction~\cite{fang2024molinstructionslargescalebiomolecularinstruction}& 0.02 & 0.35 & 41.40 & 1.00 & 0.41 & 0.23 & 0.15 \\
        \midrule
        \rowcolor{lightgray} \multicolumn{8}{c}{\textit{LLM-based generalist models}} \\

        GPT-4o~\cite{openai2024gpt4ocard} & 0.01 & 0.57 & 52.85 & 0.91 & 0.71 & 0.54 & 0.38 \\
        Deepseek-R1~\cite{deepseekai2025deepseekr1incentivizingreasoningcapability} & 0.02 & 0.56 & 92.29 & 0.57 & 0.48 & 0.38  & 0.31 \\
        Llama3-70b~\cite{llama3modelcard} & 0.03 & 0.57 & 46.63 & 0.78 & 0.57 & 0.40 & 0.30 \\
        Llama3-8b~\cite{llama3modelcard} & 0.01 &0.41 &155.17 &0.45 &0.27 &0.15 &0.11\\
        ChemDFM-13B~\cite{zhao2024chemdfmlargelanguagefoundation}& 0.32 & 0.85 & \underline{11.58} & 0.94 & 0.81 & 0.73 & 0.67 \\
        \midrule
        \rowcolor{lightgray} \multicolumn{8}{c}{\textit{Tool-based Agent models}} \\
        Agent (1-tool, Name2SMILES) & 0.25 & 0.70 & 84.81 & 0.72 & 0.67 & 0.61 & 0.56 \\
        Agent (1-tool, ChemDFM) & 0.35 & 0.86 & 12.66 & \underline{0.97} & 0.88 & 0.79 & 0.73 \\
        Agent (2-tool) & \underline{0.34} &\underline{0.87} & 12.63 & 0.94 & 0.85 & 0.80 & 0.74 \\
        Ours (Stacking Agent) & \textbf{0.38} & \textbf{0.93} & \textbf{8.68} & 0.96 & \textbf{0.92} & \textbf{0.87} &\textbf{ 0.80} \\
        \bottomrule
    \end{tabular}
    }
    \caption{Benchmark results of different models in text-based molecule design tasks. All LLM-based generalist models are evaluated on 0-shot.}
    \label{tab:text_based}
\end{table}

From the results in Tab.~\ref{tab:text_based}, our Stacking Agent outperforms other models in this task. Not only do the generated molecular structures meet the target requirements, but they also maintain high molecular validity. In contrast, models based on large language models perform poorly, particularly in terms of Exact and BLEU scores, suggesting difficulty in accurately generating target molecules in a 0-shot setting. This highlights the challenge of capturing molecular structural information without external tool support. The chemistry-specific ChemDFM performs well in the 0-shot setting, surpassing other task-specific models. However, our Stacking Agent, by combining ChemDFM with the Name2SMILES tool, achieves the best results across multiple key metrics, with a BLEU score of 0.93, outperforming ChemDFM’s 0.85.

\subsubsection{Molecule captioning}
To evaluate the ability of the model to translate complex chemical information into human-readable language descriptions, we introduce the Molecule Captioning task~\cite{guo2023largelanguagemodelschemistry}. This task requires LLMs not only to accurately recognize the molecule represented by a given SMILES string, but also to generate a concise natural language description of the molecule. 
To assess the model's performance on this task, we employ traditional natural language processing evaluation metrics, such as BLEU and ROUGE, to measure the similarity between the molecule descriptions generated by the model and the reference descriptions in the test set.
\setlength\tabcolsep{1pt}
\begin{table}[!htb]
    \centering
    \tiny
    \definecolor{lightgray}{gray}{0.9} 
    \resizebox{0.5\textwidth}{!}{
    \begin{tabular}{lccccc}
        \toprule
        \textbf{Model} & \textbf{BLEU-2↑} & \textbf{BLEU-4↑} & \textbf{ROUGE-1↑} & \textbf{ROUGE-2↑} & \textbf{ROUGE-L↑} \\
        \midrule
        \rowcolor{lightgray}\multicolumn{6}{c}{\textit{Task-specific specialist models}} \\
        Text+Chem T5~\cite{textchemt5} & 0.63 & 0.54 & \underline{0.68} & \underline{0.54} & \underline{0.62} \\
        MolXPT~\cite{liu-etal-2023-molxpt} & 0.59 & 0.50 & 0.66 & 0.51 & 0.60 \\
        InstructMol~\cite{cao2024instructmolmultimodalintegrationbuilding} & 0.48 & 0.37 & 0.57 & 0.39 & 0.50 \\
        Mol-Instruction~\cite{fang2024molinstructionslargescalebiomolecularinstruction} & 0.25 & 0.17 & 0.33 & 0.29 & 0.27 \\
        \midrule
        \rowcolor{lightgray}\multicolumn{6}{c}{\textit{LLM-based generalist models}} \\
        GPT-4o~\cite{openai2024gpt4ocard} & 0.26 & 0.17 & 0.10 & 0.00 & 0.30 \\
        Deepseek-R1~\cite{deepseekai2025deepseekr1incentivizingreasoningcapability} & 0.40 &0.25 & 0.10 &0.02 &0.21\\
        Llama3-70b~\cite{llama3modelcard} & 0.11 & 0.07 & 0.06 &0.00 & 0.12\\
        Llama3-8b~\cite{llama3modelcard} & 0.04 & 0.03 & 0.05 & 0.00 & 0.09\\
        ChemDFM-13b~\cite{zhao2024chemdfmlargelanguagefoundation}& 0.32 & 0.27 & 0.49 & 0.37 & 0.48 \\
        \midrule
        \rowcolor{lightgray}\multicolumn{6}{c}{\textit{Tool-based Agent models}} \\
        Agent (1-tool, SMILES2Description) & 0.59 & 0.52 & 0.43 & 0.29 & 0.51 \\
        Agent (1-tool, Text+ChemT5) & 0.60 &0.49 &0.39 &0.23 & 0.50\\
        Agent (2-tool) & \underline{0.64} & \underline{0.56} & 0.45 & 0.29 & 0.55 \\
        Ours (Stacking Agent) & \textbf{0.73} & \textbf{0.69} & \textbf{0.70} & \textbf{0.58} & \textbf{0.76} \\
        \bottomrule
        \end{tabular}
    }
    \caption{Benchmark results of different models in molecule captioning tasks. All LLM-based generalist models are evaluated on 0-shot.}
    \label{tab:captioning}
\end{table}

As shown in Tab.~\ref{tab:captioning}, Stacking Agent, built using the more efficient ChemHTS method, achieves the best overall performance across all metrics. For text generation tasks, task-specific models still hold a significant advantage in molecule captioning, while , aside from large parameters models like GPT-4o and Deepseek-R1, other large language models still perform poorly. Additionally, a comparison between Agent (1-tool) and Agent (2-tool) shows that increasing the number of tools boosts performance (e.g., BLEU-4 improves from 0.54 to 0.56). However, simply adding more tools doesn’t lead to major gains.

\subsubsection{Molecular Property Prediction}
Molecular property prediction~\cite{Guo_2021,wang2021chemicalreactionawaremoleculerepresentationlearning} is a fundamental task in computational chemistry that has garnered significant attention in recent years due to its potential applications in drug discovery, materials science, and other areas of chemistry. 
The task requires models to predict the chemical and physical properties of a given molecule based on its structure. The dataset consists of five tasks drawn from MoleculeNet~\cite{wu2018moleculenetbenchmarkmolecularmachine}, including BACE, BBBP, HIV, ClinTox, and Tox21. Among them, BACE and BBBP each consist of a balanced binary classification task. HIV includes an imbalanced binary classification task. ClinTox and Tox21 contain two and twenty-one imbalanced binary classification tasks, respectively. To address the severe label imbalance in these tasks, our study employs the Area Under the Receiver Operating Characteristic Curve (AUC-ROC) as the primary evaluation metric~\cite{tafvizi2022attributingaucrocanalyzebinary}. However, considering the computational challenges associated with large language models and the balanced binary task, we also incorporate Accuracy as an additional metric to provide a more comprehensive assessment.

\setlength\tabcolsep{1pt}
\begin{table}[!htb]
    \centering
    \tiny
    \definecolor{lightgray}{gray}{0.9} 
    \resizebox{0.49\textwidth}{!}{
    \begin{tabular}{lcccccccccccc}
    \toprule
\multirow{2}{*}{\textbf{Model}}            & \multicolumn{2}{c}{\textbf{BACE}} & \multicolumn{2}{c}{\textbf{BBBP}} & \multicolumn{2}{c}{\textbf{ClinTox}} & \multicolumn{2}{c}{\textbf{HIV}} & \multicolumn{2}{c}{\textbf{Tox21}} & \multicolumn{2}{c}{\textbf{Avg}} \\\cmidrule(lr){2-3} \cmidrule(lr){4-5} \cmidrule(lr){6-7} \cmidrule(lr){8-9} \cmidrule(lr){10-11} \cmidrule(lr){12-13}
                                  & \textbf{ACC↑}       & \textbf{AUC↓}      & \textbf{ACC↑}       & \textbf{AUC↓}      & \textbf{ACC↑}         & \textbf{AUC↓}       & \textbf{ACC↓}       & \textbf{AUC↑}     & \textbf{ACC↑}       & \textbf{AUC↑}      & \textbf{ACC↑}       & \textbf{AUC↓}     \\
    \midrule
    \rowcolor{lightgray}\multicolumn{13}{c}{\textit{Task-specific specialist models}} \\
Uni-Mol-v2~\cite{ji2024unimol2exploringmolecularpretraining}                        & 0.75      & \textbf{88.9}        & 0.58      & \textbf{82.6}         & \underline{0.51}        & \underline{85.3}          & \textbf{0.96}      & 90.7        & 0.92       & 80.0         & 0.74      & \textbf{85.5}        \\
MolXPT~\cite{liu-etal-2023-molxpt}                            & -         & \underline{88.4}         & -         & 80.0         & -           & \textbf{95.3}          & -         & 78.1        & -          & 77.1         & -         & 83.8        \\
InstructMol~\cite{cao2024instructmolmultimodalintegrationbuilding}                       & -         & 85.9         & -         & 64.0         & -           & -             & -         & 74.0        & -          & -            & -         & -           \\    \midrule
    \rowcolor{lightgray}\multicolumn{13}{c}{\textit{LLM-based generalist models}} \\
GPT-4o~\cite{openai2024gpt4ocard}                            & 0.38      & 38.5         & 0.56      & 57.0         & 0.51        & 51.8          & 0.59      & 54.7        & 0.37       & 36.0         & 0.48      & 47.6        \\
Deepseek-R1~\cite{deepseekai2025deepseekr1incentivizingreasoningcapability}                       & 0.62      & 52.7         & \underline{0.61}      & 63.6         & 0.48        & 48.2          & 0.51      & 50.5        & 0.75       & 60.1         & 0.59      & 55.0        \\
Llama3-70B~\cite{llama3modelcard}                        & 0.55      & 50.9         & 0.59      & 60.1         & 0.48        & 48.8          & 0.20      & 58.3        & 0.59       & 44.9         & 0.48      & 52.6        \\
Llama3-8B~\cite{llama3modelcard}                         & 0.50      & 43.8         & 0.54      & 51.4         & 0.49        & 49.5          & 0.05      & 50.5        & 0.43       & 56.1         & 0.40      & 50.2        \\
ChemDFM-13B~\cite{zhao2024chemdfmlargelanguagefoundation}                       & 0.66      & 78.4         & 0.57      & 66.7         & 0.49        & 89.9          & 0.94      & 73.6        & 0.83       & 79.8         & 0.70      & 77.7        \\
    \midrule
    \rowcolor{lightgray}\multicolumn{13}{c}{\textit{Tool-based Agent models}} \\
Agent (1-tool , SMILES2Property ) & 0.56      & 55.6         & 0.58      & 60.0         & 0.47        & 47.8          & 0.94      & 60.9        & 0.91       & 70.4         & 0.69      & 59.0        \\
Agent (1-tool , UniMol-v2 )       & 0.75      & 78.6         & 0.54      & 50.0         & 0.49        & 50.0          & 0.94      & \underline{96.8}        & 0.93       & 89.6         & 0.72      & 72.4        \\
Agent (2-tool)                    & \underline{0.75}      & 74.2         & 0.59      & 58.7         & 0.49        & 50.0          & 0.92      & 95.8        & \underline{0.94}       & \underline{91.7}         & \underline{0.74}      & 74.1        \\
Ours (Stacking Agent)             & \textbf{0.79}      & 81.4         & \textbf{0.68}      & \underline{71.1}         & \textbf{0.67}        & 72.3          & \underline{0.95}      & \textbf{97.4}        & \textbf{0.96}       & \textbf{97.8}         & \textbf{0.81}      & \underline{84.0}       
\\
    \bottomrule
    \end{tabular}
    }
    \caption{Benchmark results of different models in molecular property prediction tasks. All LLM-based generalist models are evaluated on 0-shot.}
    \label{tab:prediction}
\end{table}


From the results in Tab.~\ref{tab:prediction}, task-specific specialist models demonstrate the best overall performance. Among the models, UniMol-v2 achieves the highest average AUC-ROC and Accuracy values across the five tasks, outperforming the other models. These findings indicate that models specifically optimized for molecular property prediction tasks are more effective at learning the relationships between chemical molecular structures and their biological activities.
Moreover, the Stacking Agent structure identified through ChemHTS outperformed LLM-based generalist models on this task, especially in the imbalanced binary classification tasks of HIV and Tox21. This suggests that tool integration methods can, to some extent, compensate for the limitations of generalist models and enhance the accuracy of prediction tasks.
Finally, while our ChemHTS method falls 1.5 AUC score behind the state-of-the-art UniMol-v2 , it surpasses its average accuracy (0.74) with a score of 0.81.
\subsubsection{Reaction Prediction}
Reaction prediction is a core task in the field of chemistry, with significant importance for drug discovery, materials science, and the development of novel synthetic pathways. Given a set of reactants, the goal of this task is to predict the most likely products formed during the chemical reaction~\cite{guo2024modeling,Schwaller_2019}.
Similarly to the results of the text-based molecule design task, we used the same metrics to measure the task performance.
\setlength\tabcolsep{1pt}
\begin{table}[!htb]
    \centering
    \tiny
    \definecolor{lightgray}{gray}{0.9} 
    \resizebox{0.5\textwidth}{!}{
    \begin{tabular}{lccccccc}
        \toprule
        \textbf{Model} & \textbf{Exact↓} & \textbf{BLEU↑} & \textbf{Dis↑} & \textbf{Validity↑} & \textbf{MACCS↑} & \textbf{RDK↑} & \textbf{Morgan↑} \\
        \midrule
        \rowcolor{lightgray} \multicolumn{8}{c}{\textit{Task-specific specialist models}} \\

        Chemformer~\cite{irwin_chemformer_2022} & \textbf{0.91} & 96.1 & \underline{1.26} & 1.00 & \underline{0.97} & \underline{0.97} & \underline{0.96} \\
Text+ChemT5~\cite{textchemt5}&0.83&96.0&7.42&0.98&0.96&0.96&0.94\\
        InstructMol~\cite{cao2024instructmolmultimodalintegrationbuilding} & 0.54 & 96.7 & 10.85 & 1.00 & 0.88 & 0.78 & 0.74 \\
        Mol-Instruction~\cite{fang2024molinstructionslargescalebiomolecularinstruction} & 0.05 & 65.4 & 27.26 & 1.00 & 0.51 & 0.31 & 0.26 \\
        
        \midrule
        \rowcolor{lightgray} \multicolumn{8}{c}{\textit{LLM-based generalist models}} \\

        GPT-4o~\cite{openai2024gpt4ocard}      & 0.01 & 65.8 & 27.24  & 0.81 & 0.54 & 0.39 & 0.33 \\
Deepseek-R1~\cite{deepseekai2025deepseekr1incentivizingreasoningcapability} & 0.10 & 76.2 & 16.04  & 0.75 & 0.60 & 0.53 & 0.48 \\
Llama3-70b~\cite{llama3modelcard}  & 0.00 & 55.2 & 282.46 & 0.85 & 0.48 & 0.35 & 0.31 \\
Llama3-8B~\cite{llama3modelcard}   & 0.00 & 37.6 & 148.15 & 0.41 & 0.18 & 0.14 & 0.11 \\
ChemDFM-13B~\cite{zhao2024chemdfmlargelanguagefoundation} & 0.39 & 80.6 & 10.38  & 0.96 & 0.77 & 0.69 & 0.65 \\
        \midrule
        \rowcolor{lightgray} \multicolumn{8}{c}{\textit{Tool-based Agent models}} \\
Agent (1-tool , SMILES2Property) & 0.05       & 43.6          & 33.17         & 0.83          & 0.40          & 0.29          & 0.27          \\
Agent (1-tool , Chemformer)      & 0.89       & 96.4          & 2.44          & 1.00          & 0.97          & 0.97          & 0.95          \\
Agent (2-tool)                   & 0.87       & \underline{97.1}    & 1.6           & 1.00          & 0.97          & 0.97          & 0.95          \\
Ours (Stacking Agent)            & \underline{0.90} & \textbf{98.4} & \textbf{0.97} & \textbf{1.00} & \textbf{0.98} & \textbf{0.98} & \textbf{0.96}
\\
        \bottomrule
    \end{tabular}
    }
    \caption{Benchmark results of different models in reaction prediction tasks. All LLM-based generalist models are evaluated on 0-shot.}
    \label{tab:reaction_prediction}
\end{table}

As shown in the Tab.~\ref{tab:reaction_prediction}, It can be observed that the Chemformer model performs exceptionally well in this task, achieving a product prediction accuracy of 0.91. It also outperforms other task-specific models across all metrics. In contrast, LLMs face significant challenges, with Deepseek-R1, despite its deep reasoning capabilities, only achieving 0.10 accuracy in product prediction. Similarly, the chemistry-specific ChemDFM struggles under 0-shot conditions. The ChemHTS model excels across all metrics, except for a slightly lower exact score (0.01), surpassing the Chemformer model on all other measures. For Agent (1-tool), relying solely on the SMILES2Property tool leads to poor performance. However, leveraging the additional information provided by the RAG tool through integration, the agent also achieves significantly better performance.
\section{Analysis}
\subsection{Does it improve performance if the agent can choose from more tools?}
To further analyze the impact of the number of tools on the performance of chemical tasks, this study investigates the effect of varying tool numbers on the text-based molecule design task. 
As shown in Tab.~\ref{tab:analysis1}, under a fixed amount of training data, increasing the number of tools from 2 to 4 results in only minor changes in BLEU-2 scores. Regarding the average BLEU-2 score, both Tool Number = 2 and Tool Number = 4 achieve a score of 0.86, while Tool Number = 3 slightly underperforms with a score of 0.85. This indicates that increasing the number of tools has limited benefits for BLEU-2 score improvement, which may be attributed to redundancy in contributions among tools and the fact that the quality of individual tools determines their impact on the final performance.

\setlength\tabcolsep{1pt}
\begin{table}[!htb]
    \centering
    \tiny
    \definecolor{lightgray}{gray}{0.9} 
    \resizebox{0.49\textwidth}{!}{
    \begin{tabular}{ccccccccccc}
        \toprule
        \multirow{2}{*}{\textbf{Tool Number}} & \multicolumn{8}{c}{\textbf{Train Data Number}} & \multirow{2}{*}{\textbf{AVG}} \\ 
        \cmidrule(lr){2-9}
        & \multicolumn{2}{c}{5} & \multicolumn{2}{c}{10} & \multicolumn{2}{c}{20} & \multicolumn{2}{c}{30} & \\ 
        \cmidrule(lr){2-3} \cmidrule(lr){4-5} \cmidrule(lr){6-7} \cmidrule(lr){8-9}
        & Layer & BLEU-2 & Layer & BLEU-2 & Layer & BLEU-2 & Layer & BLEU-2 & \\
        \midrule
        2 & 0.6 & 0.79 & 2.7 & 0.89 & 3.0 & 0.89 & 3.2 & 0.87 & 0.86 \\
        3 & 0.4 & 0.81 & 2.5 & 0.87 & 2.8 & 0.86 & 3.0 & 0.85 & 0.85 \\
        4 & 0.6 & 0.80 & 2.8 & 0.88 & 3.0 & 0.87 & 3.2 & 0.86 & 0.86 \\
        \bottomrule
    \end{tabular}
    }
    \caption{Comparison of the performance of different tools (Tool Layerber) on the text-based molecule design task under different training data sizes (Train Data Layerber). And Layer represents the number of layers of stacking agents.}
    \label{tab:analysis1}
\end{table}

\subsection{Does more training data lead to better performance?}
Our study also conducts an in-depth analysis of the performance differences in the text-based molecule design task among the optimal stacking structures identified by the ChemHTS method under varying amounts of training data.
As shown in Tab.~\ref{tab:analysis1}, BLEU-2 scores generally exhibit an upward trend with the increase in training data (from 5 to 30). For example, for Tool Number = 2, the BLEU-2 score is 0.79 when the training data is 5, and it improves to 0.87 when the training data increases to 30. Similar trends are observed for other tool numbers, indicating that increasing the amount of training data enhances the model's translation quality. This trend aligns with expectations; however, more training data does not always guarantee better performance. Therefore, further experimental analysis is required to determine the optimal amount of training data for different tasks and select the best combination.

\setlength\tabcolsep{1pt}
\begin{table}[!htb]
    \centering
    \tiny
    \definecolor{lightgray}{gray}{0.9} 
    \resizebox{0.4\textwidth}{!}{
    \begin{tabular}{lccccccccc}
    \toprule
    \textbf{Layer} & 1 & 2 & 3 & 4 & 7 & 8 & 10 & 10+ & \textbf{AVG} \\
    \midrule
    \textbf{BLEU-2} & 0.88 & 0.90 & 0.92 & 0.93 & 0.90 & 0.92 & 0.91 & 0.90 & 0.91 \\
    \bottomrule
    \end{tabular}
    }
    \caption{Comparison of different tool stacking levels on the performance of text-based molecule design tasks.}
    \label{tab:analysis2}
\end{table}

\subsection{Does increasing the number of stacking layers improve performance?}

This section uses the text-based molecule design task as a representative to explore whether increasing the number of stacking layers can improve performance.  
As shown in the results from Tab.~\ref{tab:analysis2}, there appears to be an upper limit to the performance improvement achieved by increasing the number of stacking layers. 
When the number of layers is small (e.g., 1 or 2 layers), the model may fail to sufficiently capture complex features, resulting in lower BLEU-2 scores. As the number of layers increases to 7, the model's learning capacity improves, and the generation quality reaches its peak. However, further increasing the number of layers does not lead to continued performance improvement. This could be attributed to the introduction of redundant information with excessive layers, causing a decline in text generation quality or overfitting to the training data.

\setlength\tabcolsep{1pt}
\begin{table*}[!htb]
    \centering
    \tiny
    \definecolor{lightgray}{gray}{0.9} 
    \resizebox{0.95\textwidth}{!}{
    \begin{tabular}[l]{|>{\raggedright\arraybackslash}m{1.5cm}|>{\raggedright\arraybackslash}m{4cm}|>{\raggedright\arraybackslash}m{10cm}|}
    \hline
    \textbf{Stacking Pattern} & \textbf{Input} & \textbf{Output} \\ 
    \hline
    Correct 
    & The molecule is a cycloalkane composed of a three-carbon ring. It functions as an inhalation anesthetic. It is a cycloalkane and a member of the cyclopropane class. What is its predicted SMILES?
    \newline
    \newline
    \newline
    \textcolor{red}{Gold Answer: C1CC1}
    & According to the provided description, the molecule is cyclopropane, which is a cycloalkane formed by three carbon atoms in a ring, well known for its use as an inhalation anesthetic. To confirm the SMILES notation of cyclopropane, I used the ChemDFM tool.
    \newline
    \textbf{Action:} ChemDFM  
    \newline
    \textbf{Input:} What is the SMILES notation for cyclopropane (a three-carbon cycloalkane used as an inhalation anesthetic)?  
    \newline
    \textbf{Output:} CC1(C)CC1  \textcolor{red}{\ding{55}}
    \newline
    \textbf{Observation:} The returned "CC1(C)CC1" does not correspond to cyclopropane. Cyclopropane should be a simple three-carbon ring, typically represented as "C1CC1". Next, I will call the Name2SMILES tool to further correct the result.
    \newline
    \textbf{Action:} Name2SMILES  
    \newline
    \textbf{Input:} Cyclopropane  
    \newline
    \textbf{Output:} C1CC1  
    \newline
    \textbf{Observation:} Based on the known structure of cyclopropane, the result from this tool is correct. Therefore, I will provide the correct SMILES based on the understanding of the chemical structure.  
    \newline
    \textbf{Final Answer: C1CC1 \textcolor{red}{\ding{52}}}\\
    \hline
    \end{tabular}
    }
    \caption{Case analysis of \textbf{"Correct"} behavior patterns of agents during tools stacking.}
    \label{tab:case study}
\end{table*}

%
\subsection{Why tool stacking works?}
\label{6.4}
This chapter explores the behavioral patterns exhibited by agents during the process of tool stacking through case analysis and examines how tool stacking enhances task performance. From the case studies, we identify \textbf{four distinct stacking behavior patterns—correct, modify, judge and reserve}, with detailed information on all cases provided in Appendix.~\ref{casestudy}. Due to space limitations, this paper illustrates only one \textbf{correct} stacking behavior pattern as shown in Tab.~\ref{tab:case study}.
In the text-based molecular design task, two callable tools are provided to the model. The experiment first invokes the ChemDFM tool for SMILES prediction, which returns "C1(C)CC1". However, this result does not conform to the standard SMILES representation of cyclopropane. The error likely arises from ChemDFM's misinterpretation of cyclic structures, as it fails to correctly identify the single three-membered ring and instead erroneously introduces an additional branch. To address this issue, the experiment subsequently invokes the Name2SMILES tool with "Cyclopropane" as the query. This tool correctly returns "C1CC1", which is confirmed through subsequent analysis to align with the known standard structure of cyclopropane. Consequently, the final SMILES output is corrected to "C1CC1", meeting the requirements of the standard answer.
This experiment shows that individual tools may have certain limitations when handling chemical structure prediction tasks, as evidenced by ChemDFM's erroneous prediction. However, by employing the tool stacking strategy, the experiment effectively leverages the complementary strengths of multiple tools, significantly improving the accuracy of the final prediction.

\subsection{Comparison with LLM-based Multi-Agent Systems}
LLM-based Multi-Agent Systems (MAS) and Tool-augmented LLM share similarities in areas such as task decomposition, tool invocation, and information sharing. Therefore, our study focuses on comparing the performance of six multi-agent systems with different communication structures against the optimal stacking agent path proposed by our ChemHTS method in the text-based molecule design task. For details on the specific multi-agent systems, please refer to Appendix.~\ref{Framework},~\ref{multi-agent results}.
\begin{figure}[t] 
    \centering
            \includegraphics[width=0.45\textwidth]{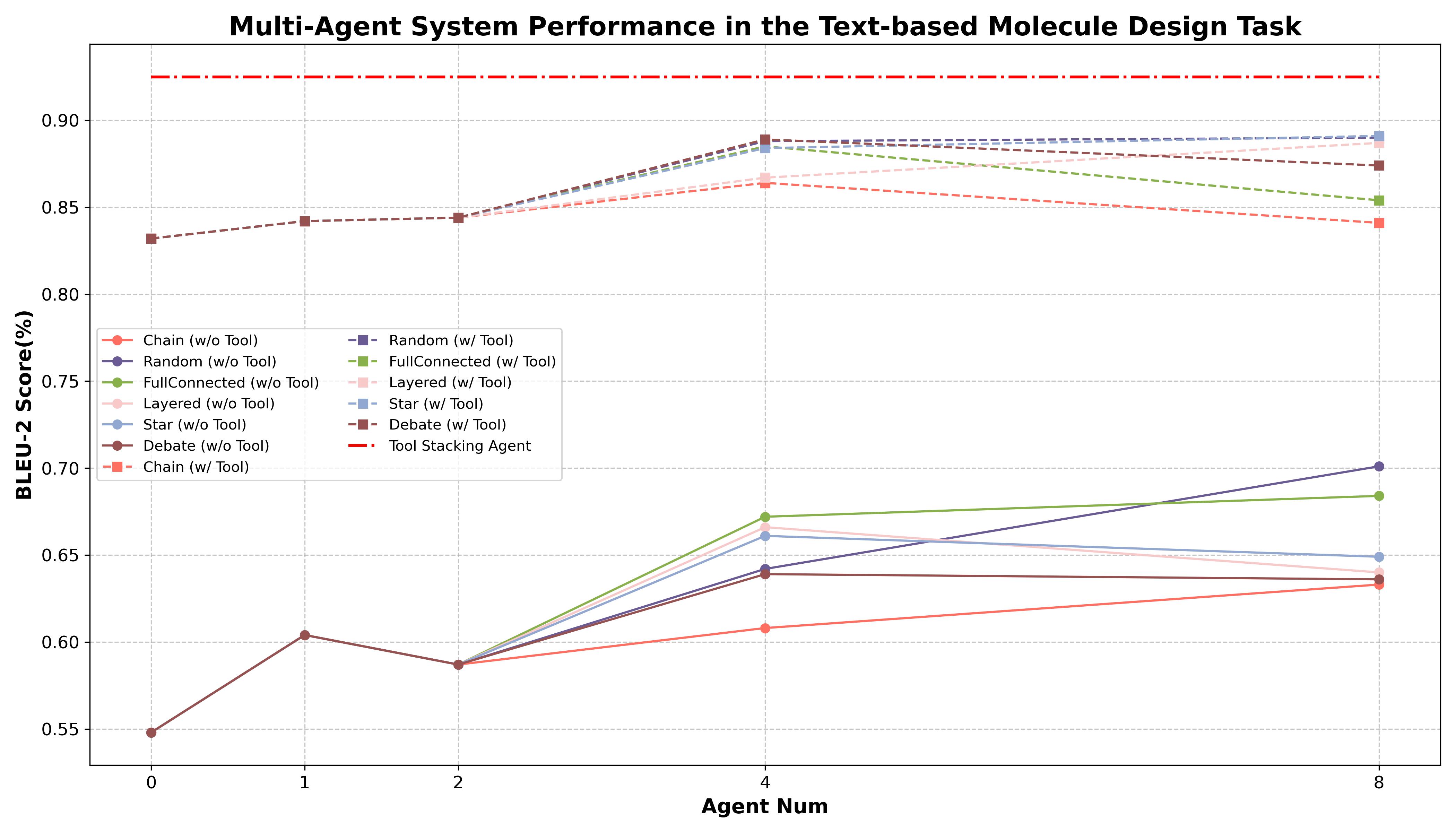}
    \caption{Performance comparison of 6 multi-agent systems with different communication structures and our optimal stacking agent path on the text-based molecule design task.
    }
    \label{fig:multiagent}
\end{figure}

As shown in Fig.~\ref{fig:multiagent}, the BLEU-2 scores of various communication structures improve as the scale of agents increases. However, the growth trends differ significantly. When the agent scale becomes larger, the performance of different structures begins to diverge. Among them, the Full-Connected and Layered structures demonstrate significantly higher quality values compared to other structures. 
Nevertheless, the performance of the optimal stacking agent path surpasses the upper performance limit of the multi-agent systems. This is because the optimal path can more effectively leverage task decomposition and tool invocation capabilities while avoiding potential communication overhead and coordination bottlenecks inherent in multi-agent systems. 
More specific experimental results of multi-agent can be found in Appendix.~\ref{appendix mulagent results}.

\section{Conclusion}

Our study proposes \textbf{ChemHTS}, a hierarchical tool stacking method to enhance tool utilization in LLM-driven chemistry tasks. By addressing key challenges such as bias from single tool calls and lack of tool collaboration, ChemHTS improves reasoning and computational efficiency.
Experiments across four chemistry tasks show that ChemHTS outperforms both general LLMs and chemistry-specific models. Additionally, we identify four tool stacking behavioral patterns—Correct, Modify, Judge, and Reserve—which enhance interpretability.
In summary, our work highlights the importance of tool collaboration in scientific AI, providing a systematic framework for optimizing LLM-agent interactions in chemistry.

\section*{Limitations}
Despite the promising results of ChemHTS in optimizing tool invocation pathways for chemical tasks, several limitations remain. First, the method relies on predefined toolsets, which may not generalize well to novel or underrepresented chemistry tasks. Expanding the adaptability of ChemHTS to dynamically incorporate new tools remains an open challenge. Second, the hierarchical stacking strategy assumes that optimal tool combinations can be effectively learned from limited training data, yet real-world chemistry problems often require extensive domain expertise, which LLMs may struggle to acquire solely through tool interactions. Finally, while ChemHTS improves tool collaboration, its computational cost increases with the number of tool-stacking layers, potentially limiting its scalability in large-scale applications. Future work should explore more efficient optimization strategies and adaptive learning mechanisms to enhance both generalizability and efficiency.

\bibliography{custom}
\clearpage

\appendix
\renewcommand{\thesection}{\Alph{section}} 
\section*{Appendix}
\begin{figure*}[t] 
    \centering            \includegraphics[width=0.90\textwidth]{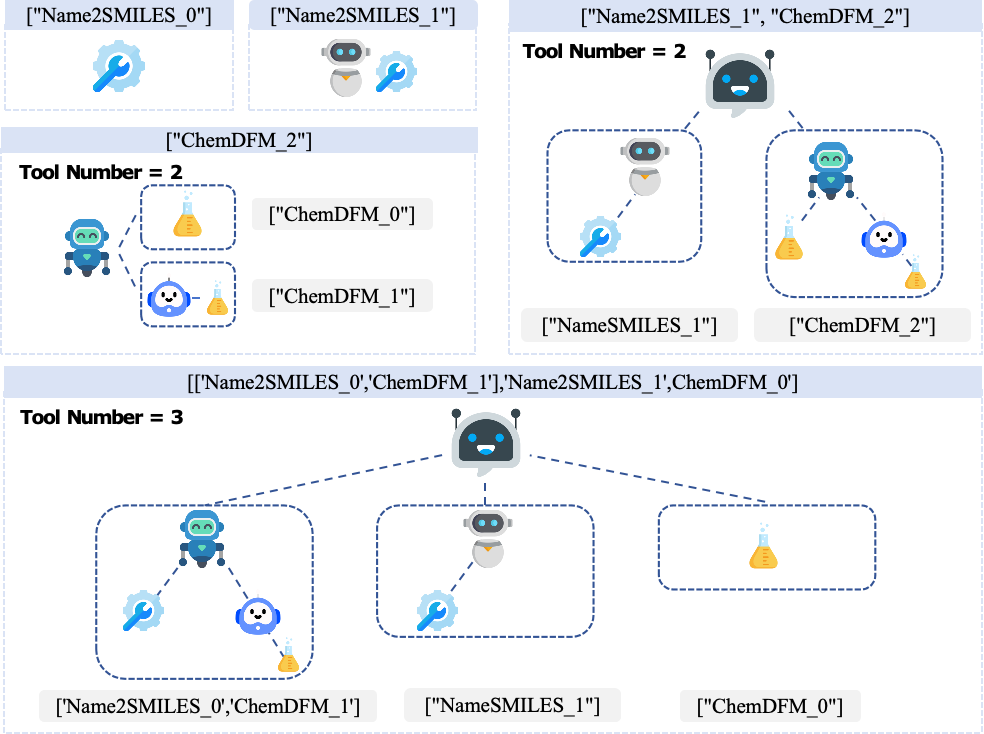}
    \captionsetup{font={small}} 
    \caption{
    The example of the Naming rule, where the icon "\includegraphics[height=1.2em]{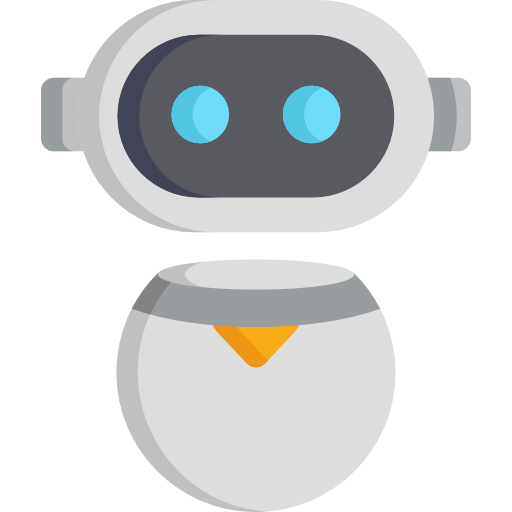}","    \includegraphics[height=1.2em]
{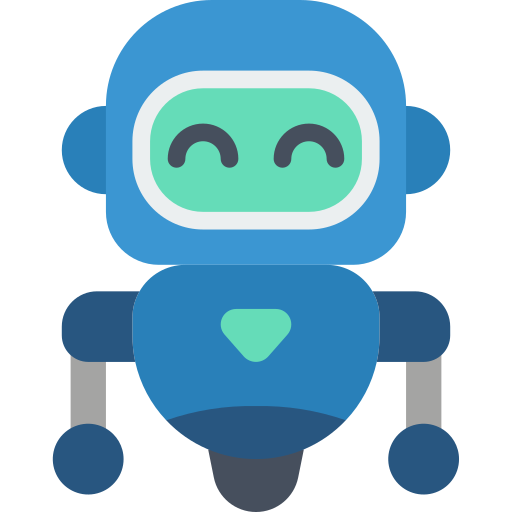}", "\includegraphics[height=1.2em]{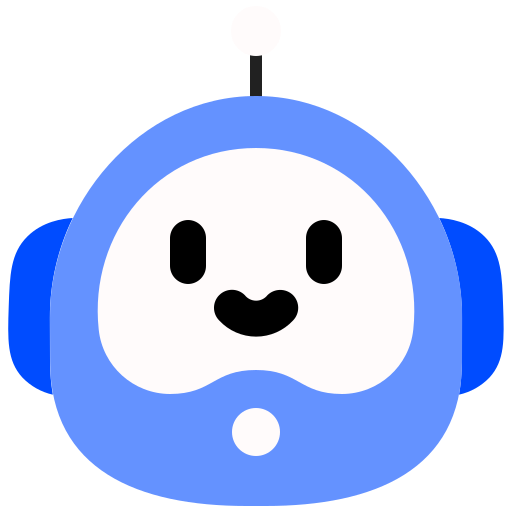}" represent the Agent and the icon '\includegraphics[height=1.2em]{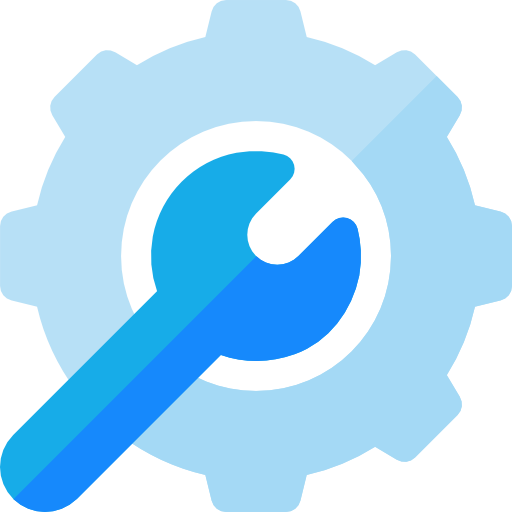}','\includegraphics[height=1.2em]{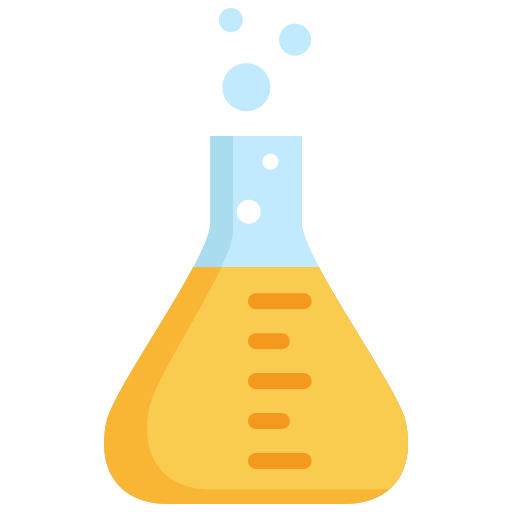}' represent the Retrieve tool and Compute tool respectively.}
    \label{fig:appendix_name}
\end{figure*}

\section{Stacking Agent Details}
\label{Stacking Agent Details}
\setlength\tabcolsep{1pt}
\begin{table*}[!htb]
    \centering
    \captionsetup{font={small}} 

    \resizebox{0.93\textwidth}{!}{
    \begin{tabular}{|c|lc|c|}
\hline
\multirow{2}{*}{Task}                                                                                  & \multicolumn{2}{c|}{Train}                                                                                                   & \multirow{2}{*}{Result} \\ \cline{2-3}
                                                                                                       & \multicolumn{1}{c|}{Final tool}                                                                                      & score &                         \\ \hline
\multirow{3}{*}{\begin{tabular}[c]{@{}c@{}}Text-based\\ Molecule\\ Design\end{tabular}}                & \multicolumn{1}{l|}{{[}'Name2SMILES\_3', 'ChemDFM\_0'{]}}                                                            & 0.80  & 0.90                    \\ \cline{2-4} 
                                                                                                       & \multicolumn{1}{l|}{\textbf{{[}{[}'ChemDFM\_0', 'Name2SMILES\_1'{]}, 'ChemDFM\_1'{]}}}                               & 0.81  & \textbf{0.93}           \\ \cline{2-4} 
                                                                                                       & \multicolumn{1}{l|}{{[}{[}'ChemDFM\_1','Name2SMILES\_1'{]},{[}'ChemDFM\_1','Name2SMILES\_2'{]}{]}}                   & 0.91  & 0.91                    \\ \hline
\multirow{3}{*}{\begin{tabular}[c]{@{}c@{}}Molecule\\ Captioning\end{tabular}}                         & \multicolumn{1}{l|}{\textbf{{[}'SMILES2Description\_2','TextChemT5\_0'{]}}}                                          & 0.79  & \textbf{0.73}           \\ \cline{2-4} 
                                                                                                       & \multicolumn{1}{l|}{{[}'SMILES2Description\_3'{]}}                                                                   & 0.70  & 0.65                    \\ \cline{2-4} 
                                                                                                       & \multicolumn{1}{l|}{{[}'TextChemT5\_1','SMILES2Description\_1'{]}}                                                   & 0.71  & 0.66                    \\ \hline
\multirow{3}{*}{\begin{tabular}[c]{@{}c@{}}Reaction\\ Prediction\end{tabular}}                         & \multicolumn{1}{l|}{\textbf{{[}'Chemformer\_0','SMILES2Property\_2'{]}}}                                             & 1.00  & \textbf{0.90}                    \\ \cline{2-4} 
                                                                                                       & \multicolumn{1}{l|}{{[}'Chemformer\_2'{]}}                                                                           & 0.90  & 0.89                    \\ \cline{2-4} 
                                                                                                       & \multicolumn{1}{l|}{{[}'Chemformer\_1','SMILES2Property\_1'{]}}                                                      & 0.90  & 0.85                    \\ \hline
\multirow{3}{*}{\begin{tabular}[c]{@{}c@{}}Property\\ Prediction\\ (BACE)\end{tabular}}    & \multicolumn{1}{l|}{{[}'UniMol\_1','SMILES2Property\_1'{]}}                                                          & 0.80  & 72.1                    \\ \cline{2-4} 
                                                                                                       & \multicolumn{1}{l|}{\textbf{{[}{[}'UniMol\_1', 'SMILES2Property\_0'{]}, 'SMILES2Property\_0'{]}}}                    & 0.82  & \textbf{81.4}                    \\ \cline{2-4} 
                                                                                                       & \multicolumn{1}{l|}{{[}'UniMol\_3'{]}}                                                                               & 0.75  & 78.6                    \\ \hline
\multirow{3}{*}{\begin{tabular}[c]{@{}c@{}}Property\\ Prediction\\ (BBBP)\end{tabular}}    & \multicolumn{1}{l|}{\textbf{{[}'UniMol\_1','SMILES2Property\_2'{]}}}                                                 & 0.68  & \textbf{71.1}           \\ \cline{2-4} 
                                                                                                       & \multicolumn{1}{l|}{{[}'SMILES2Property\_3'{]}}                                                                      & 0.75  & 69.3                    \\ \cline{2-4} 
                                                                                                       & \multicolumn{1}{l|}{{[}'UniMol\_2'{]}}                                                                               & 0.73  & 70.6                    \\ \hline
\multirow{3}{*}{\begin{tabular}[c]{@{}c@{}}Property\\ Prediction\\ (Clintox)\end{tabular}} & \multicolumn{1}{l|}{\textbf{{[}{[}'UniMol\_1', 'SMILES2Property\_0'{]}, {[}'UniMol\_1', 'SMILES2Property\_1'{]}{]}}} & 0.70  & \textbf{72.3}           \\ \cline{2-4} 
                                                                                                       & \multicolumn{1}{l|}{{[}'SMILES2Property\_2'{]}}                                                                      & 0.68  & 61.4                    \\ \cline{2-4} 
                                                                                                       & \multicolumn{1}{l|}{{[}'UniMol\_1','SMILES2Property\_1'{]}}                                                          & 0.65  & 69.1                    \\ \hline
\multirow{3}{*}{\begin{tabular}[c]{@{}c@{}}Property\\ Prediction\\ (HIV)\end{tabular}}     & \multicolumn{1}{l|}{\textbf{{[}{[}'SMILES2Property\_1', 'UniMol\_0'{]}, 'SMILES2Property\_1'{]}}}                    & 1.00  & \textbf{97.4}           \\ \cline{2-4} 
                                                                                                       & \multicolumn{1}{l|}{{[}'UniMol\_0', 'SMILES2Property\_1'{]}}                                                         & 0.85  & 90.1                    \\ \cline{2-4} 
                                                                                                       & \multicolumn{1}{l|}{{[}'UniMol\_2'{]}}                                                                               & 0.90  & 96.9                    \\ \hline
\multirow{3}{*}{\begin{tabular}[c]{@{}c@{}}Property\\ Prediction\\ (Tox21)\end{tabular}}   & \multicolumn{1}{l|}{\textbf{{[}'UniMol\_2'{]}}}                                                                      & 0.78  & \textbf{92.3}           \\ \cline{2-4} 
                                                                                                       & \multicolumn{1}{l|}{{[}'UniMol\_0','SMILES2Property\_2'{]}}                                                          & 0.80  & 79.6                    \\ \cline{2-4} 
                                                                                                       & \multicolumn{1}{l|}{{[}'SMILES2Property\_2'{]}}                                                                      & 0.85  & 74.9                    \\ \hline
\end{tabular}
}
    
    \caption{Stacking results of different tasks. The \textbf{bold} font represents the most suitable toolset obtained in the task experiment.}
    \label{tab:appendix_molecule_design}
\end{table*}

We conducted several experiments and selected three optimal hierarchical stacking toolsets, and now we will present the stacking results and scores for each stacking agent in the Tab.~\ref{tab:appendix_molecule_design}, along with the corresponding prompts:
\subsection{Stacking Agent}
\subsubsection{Prompt}
Our Agent framework is based on the ReAct method~\cite{yao2023reactsynergizingreasoningacting} to implement tool  and reasoning processes. 
\subsubsection{Naming Rule}
To facilitate the comprehension of Hierarchical-Tool-Stacking, we propose a systematic hierarchical naming rules as follows in Fig.~\ref{fig:appendix_name}. In addition, in order to prevent the tool name from affecting the agent call, we choose to anonymously process the agent tool, that is, \{task name\}\_\{num\}.
\begin{itemize}
    \item \textbf{Self-Stacking Tools}: Hierarchical proxies are constructed through a recursive generation strategy, with the depth of the hierarchy dynamically extendable via the numerical suffix. For instance in Fig.~\ref{fig:appendix_name}, "["Name2SMILES\_0"]" denotes the base tool, while "["Name2SMILES\_1"]" signifies a first-layer tool, referred to as an Agent Tool (which encapsulates both the tool and the Agent into a new tool) and "["ChemDFM\_2"]".

    \item \textbf{Multiple Tool Combinations}: The combination of multiple tools within an Agent is represented in a list format, utilizing depth-first traversal to generate sub-tools, thereby forming the final toolset for the agent. For example in Fig.~\ref{fig:appendix_name}, the structure "["Name2SMILES\_1", "ChemDFM\_2"]" represents a flat structure with tools at the same level ([A, B, ...]), while the structure "[['Name2SMILES\_0','ChemDFM\_1'],'Nam
    e2SMILES\_1','ChemDFM\_0']" illustrates a nested structure ([[A, B], C, D]), where tools A and B are first combined before being integrated with tool C and D.
\end{itemize}


\subsection{Text-based Molecule Design}
\subsubsection{Task Introduction}
The test set of ChEBI-20-MM is exploited for this task in ChemLLMBench. Models are asked to predict the SMILES of the molecule that fits the given description. Considering the low accuracy of the models, we use BLEU-2 as the training metric and use metrics such as Exact, Dis and others during the testing stage.
\subsubsection{Prompt}
We use a simpler prompt compared with the prompt introduced in \cite{guo2023largelanguagemodelschemistry}
\begin{tcolorbox}[colback=gray!10, colframe=black, title=Prompt: Text-based Molecule Design]
You are an expert chemist. Given the molecular requirements description, your task is to design a new molecule SMILES:\\
Molecular requirements description::
\end{tcolorbox}
\subsection{Molecule Captioning}
\subsubsection{Task Introduction}
The test set  is the same with the Text-based Molecule Design task. Because this is the mirroring task, which generates a detailed description by giving a SMILES to the models. In this task, we also choose the BLEU-2 as the metric in the training stage. When in the test stage, more metrics, like BLEU and ROUGE, are utilized to Measure the performance of the model.
\subsubsection{Prompt}
We also use a simpler prompt compared with the prompt introduced in \cite{guo2023largelanguagemodelschemistry}
\begin{tcolorbox}[colback=gray!10, colframe=black, title=Prompt: Molecule Captioning]
You are an expert chemist. Given the molecular SMILES, your task is to provide the detailed description((The molecule is ...) of the molecule. \\Please strictly follow the format, no other information can be provided.\\Molecular SMILES:
\end{tcolorbox}

\subsection{Molecular Property Prediction}
\subsubsection{Task Introduction}
The molecular property prediction tasks in ChemLLMBench consist of five tasks from MoleculeNet benchmark~\cite{wu2018moleculenetbenchmarkmolecularmachine}, including BACE, BBBP, HIV, ClinTox, and Tox21. Among these, BACE and BBBP are balanced binary classification tasks, while HIV represents an unbalanced binary classification task. ClinTox consists of 2 unbalanced binary classification tasks, and Tox21 comprises 21 unbalanced binary classification tasks. In this task, we choose the AUC-ROC as the first metric in the training stage. Considering the calculation method of AUC-ROC for large language models, we also introduced Accuracy as a second metric for test stage.
\subsubsection{Prompt}
We use the same prompts introduced in ~\cite{guo2023largelanguagemodelschemistry}

\subsection{Reaction Prediction}
\subsubsection{Task Introduction}
The reaction prediction task asks the model to predict the product of the given reaction. ChemLLMBench utilizes the USPTO-MIT dataset for this task. Since the benchmark metric is Accuracy, we also chose Accuracy as the training metric, and considering that the answer is also SMILES, we adopted the same metric as Molecular Design task for measurement during the testing stage.
\subsubsection{Prompt}
We reformat the prompt provided ~\cite{guo2023largelanguagemodelschemistry}.
\begin{tcolorbox}[colback=gray!10, colframe=black, title=Prompt: Reaction Prediction]
Given an incomplete chemical reaction equation in SMILES notation (format: reactants>>product, where multiple reactants are separated by dots '.'), predict and complete the missing products marked as '\_\_\_'. The response should only contain the only one SMILES representation of the missing molecule, without any additional explanation (Note: Please only output only one final product). Please answer the question based on the following Chemical reaction equation:
\end{tcolorbox}

\section{Multi-agent Implementation Details}

In this section, we will discuss how to implement multi-agent systems and specific ways of information transmission, including chain, random, star, full-connected, layered, and debate graphs.

\subsection{Framework}
\label{Framework}
In order to complete chemical tasks, we divided multi-agent into two types: agents with tools and agents without tools, and tested them on the first task, the Text-based Molecule Design task. The overall of our multi-agent framework is a modification of the framework of ~\cite{zhang2024cutcrapeconomicalcommunication} and ~\cite{qian2024scalinglargelanguagemodelbasedmultiagentcollaboration} that utilized different spatial and temporal masks to complete in the following six multi-agent structures: Chain, Random, FullConnected, Layered, Star and Debate mode. \\
However, since we are modifying their approach with a greater focus on information transmission and are also limited by API calls, we can only make a one-sided comparison regarding the recording of tokens and time.
\subsection{Implementation Details}
In multi-agent systems, information transmission is a critical factor for enhancing performance. By utilizing various structures, information can be conveyed through multiple pathways. To improve the efficiency of information collection, we have adopted the Final decision approach. Specifically, at the end of all structures, we have integrated a FinalRefer Agent to perform the final summary and decision-making. The FinalRefer prompt is followed:
\begin{tcolorbox}[colback=gray!10, colframe=black, title=Prompt: FinalRefer]
You are a strategic planning and final integration agent. You will be given a graduate-level question and reasoning outputs from all other agents.
Your task is to integrate all the information into a single, cohesive answer with detailed reasoning and evidence.\\
Your final output should:
1. Summarize the contributions from all agents, highlighting key insights.\\
3. Provide the final answer with a clear and detailed explanation.\\
4. Conclude with the final answer on a new line with the format: "The final answer is 'SMILES'\\
Here is the question:{question}. At the same time, the output of other agents is as follows:\\
{answers}
\end{tcolorbox}

In the implementation of the agents with tools, we modified all agents along the path except for the Final agent, while still following the ReAct framework for tool calling. During this process, due to the constraints of API calls, both the time required and the number of tokens used will be greater compared to agents without tools.
\subsection{Spatial Communication Topologies}
\subsubsection{Chain}
\begin{figure}[!h] 
    \centering
    \includegraphics[width=0.35\textwidth]{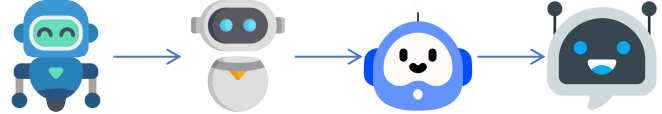}
    \captionsetup{font={small}} 
    \caption{
    Demonstration of \textbf{chain} structure}
    \label{fig:chain}
\end{figure}

The chain graph (Fig.~\ref{fig:chain}) is one of the most widely utilized communication architectures in contemporary multi-agent systems. In this architecture, the first agent receives input from the user, transforms it into new instruction, and subsequently forwards it to the next agent. Generally, the final agent in the chain provides a summary and answers.
\subsubsection{Random}
\begin{figure}[!h] 
    \centering
    \includegraphics[width=0.19\textwidth]{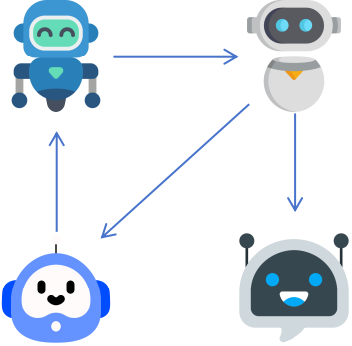}
    \captionsetup{font={small}} 
    \caption{
    Demonstration of \textbf{random} structure}
    \label{fig:random}
\end{figure}

The random graph refers to a sparse graph randomly sampled from a complete graph, as shown in the Fig.~\ref{fig:random}. They will execute asynchronously in multiple rounds and then randomly transmit information to the target agent. Finally, all the answers and information will transmit to the Final agent to make a final answer.
\subsubsection{FullConnected}
\begin{figure}[!h] 
    \centering
    \includegraphics[width=0.29\textwidth]{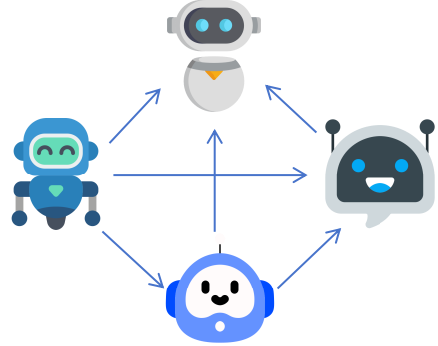}
    \captionsetup{font={small}} 
    \caption{
    Demonstration of \textbf{FullConnected} structure}
    \label{fig:fullconnected}
\end{figure}

The fullconnected graph (Fig.~\ref{fig:fullconnected}) is a directed graphs compared to traditional fully linked undirected graphs, which transmit information in a certain order to complete this topology structure. The final agent summarizes the dialogue and provides a concluding output or reflection.
\subsubsection{Layered}
\begin{figure}[!h] 
    \centering
    \includegraphics[width=0.35\textwidth]{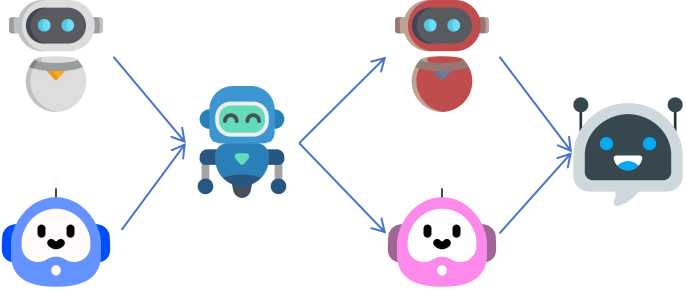}
    \captionsetup{font={small}} 
    \caption{
    Demonstration of \textbf{Layered} structure}
    \label{fig:Layered}
\end{figure}

The layered graph (Fig.~\ref{fig:Layered},~\cite{qian2024scalinglargelanguagemodelbasedmultiagentcollaboration}) refers to a stacked configuration similar to a multilayer perceptron (MLP). The first layer agents will feed to the agents in the second layer, and the final layer will make the summary and final-decision.
\subsubsection{Star}
\begin{figure}[!h] 
    \centering
    \includegraphics[width=0.23\textwidth]{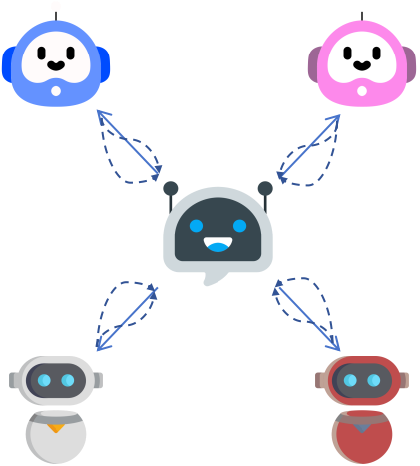}
    \captionsetup{font={small}} 
    \caption{
    Demonstration of \textbf{Star} structure}
    \label{fig:Star}
\end{figure}

The star grpah (Fig.~\ref{fig:Star}) resembles the tree structure. Firstly, the problem will be handed over to the external leaf nodes for processing, and the obtained answer will be passed to the central root node, which will be repeated multiple times. Finally, the root node will give a summary and make the descision.
\subsubsection{Debate}
\begin{figure}[!h] 
    \centering
    \includegraphics[width=0.29\textwidth]{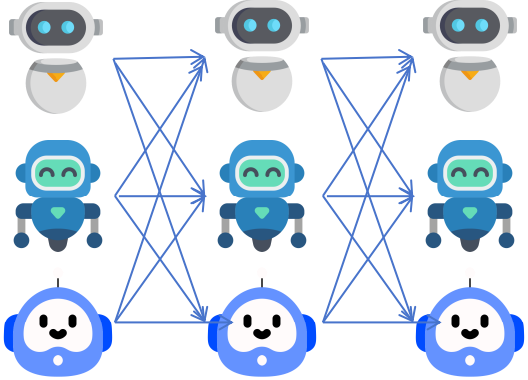}
    \captionsetup{font={small}} 
    \caption{
    Demonstration of \textbf{Debate} structure}
    \label{fig:Debate}
\end{figure}

The debate graph (Fig.~\ref{fig:Debate},~\cite{du2023improvingfactualityreasoninglanguage}) is multiple agents to engage in a debate, where in each round, every agent receives the outputs of all agents from the previous round before making their own statements. Generally, the finalRefer agent will help them to make the final descision.

\subsubsection{Tree-like\&Ours}
\begin{figure}[!h] 
    \centering
    \includegraphics[width=0.5\textwidth]{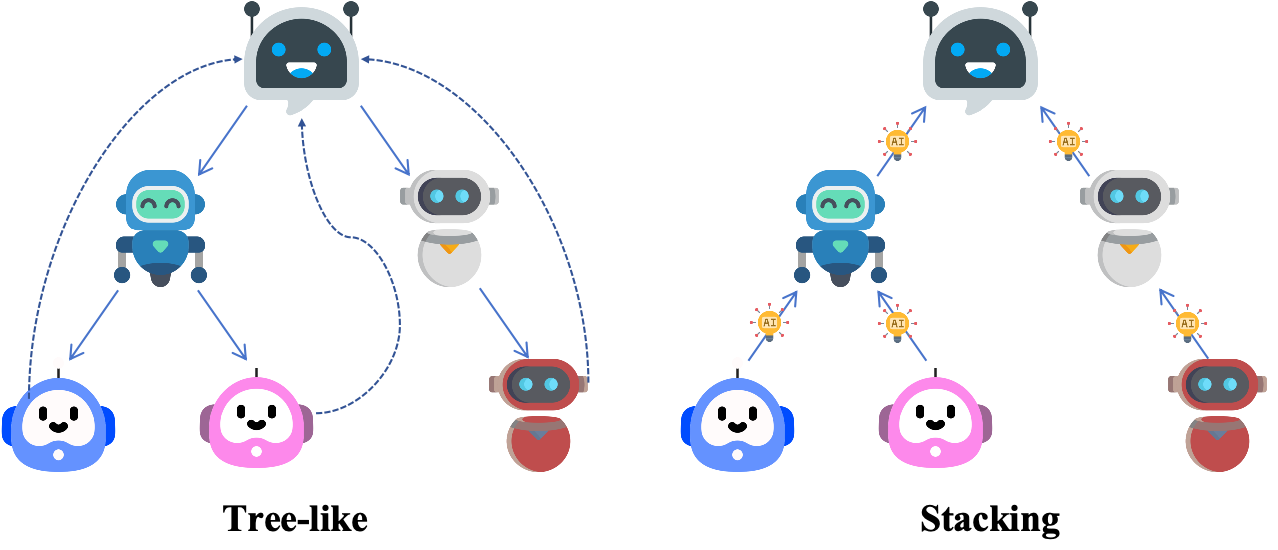}
    \captionsetup{font={small}} 
    \caption{
    Demonstration of \textbf{Tree} and \textbf{Ours} structure. In the stacking structure, the icon '\includegraphics[height=1em]{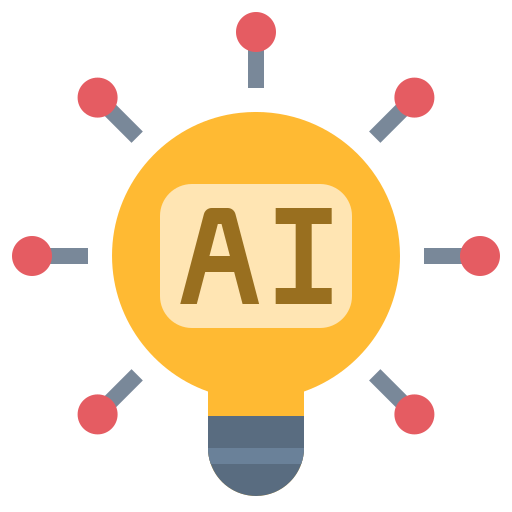}' means the parent node can choose which child node's information to accept.}
    \label{fig:Stacking}
\end{figure}

In order to better understand our stacking structure, we will compare it with a tree like multi-agent pipeline that is relatively similar. The tree graph usually has the root node as the manager to supervise the nodes below to complete various tasks, and finally return the results of the leaf nodes to the root node for processing. Overall, this is just a top-down process of information transmission. As showin in the Fig.~\ref{fig:Stacking}\\
As for our stacking structure, from the perspective of information transmission, the main agent of the root node also receives the information completed from below, but there is a difference. For the so-called tool node, it is a bottom-up process. After continuous information superposition step by step, it is passed upward through the parent node and then given to the main agent for processing. It can be seen from the figure that in each transmission process, whether it is the root node or the parent node, they all selectively accept the information from the child node, and it is not like a tree structure that is passed downward.

\section{Multi-agent Experimental Results}
\label{multi-agent results}
The experimental results are shown in the Tab.~\ref{tab:multiagent}. From the table, it can be observed that for different structures, there is an initial performance improvement as the value of num increases. However, when num reaches 8, only the Random, Layered, and Star modes show some improvement, with a maximum score of 0.891. In contrast, our best combined structure ([[‘ChemDFM\_0’, ‘Name2SMILES\_1’], ‘ChemDFM\_1’]) achieves a score of \textbf{0.925} with a num of only 4, indicating a significant enhancement.\\
Furthermore, since our multi-agent framework has been modified from others, the number of rounds for information transmission has not been optimized, leading to a substantial increase in both token count and time. This also suggests that even in complex and prolonged reasoning scenarios, relying solely on information transmission does not yield particularly high improvements. 
\setlength\tabcolsep{1pt}
\begin{table}[!htb]
    \centering
    \tiny
    \definecolor{lightgray}{gray}{0.9} 
    \resizebox{0.5\textwidth}{!}{
\begin{tabular}{cccccccccc}
\toprule
\rowcolor{lightgray} \multicolumn{10}{c}{Text-based Molecule Design (BLEU-2) - w/o-Tool} \\ 
\midrule
NUM  & Chain  & Random  & FullConnected  & Layered  & Star   & Debate  & Avg   & Avg\_all\_tokens & Avg\_Time \\ 
\midrule
0    & 0.548  & 0.548   & 0.548          & 0.548    & 0.548  & 0.548   & 0.548 & 890.19           & 8.254     \\
1    & 0.604  & 0.604   & 0.604          & 0.604    & 0.604  & 0.604   & 0.604 & 2315.22          & 15.826    \\
2    & 0.587  & 0.587   & 0.587          & 0.587    & 0.587  & 0.587   & 0.587 & 3619.54          & 21.034    \\
4    & 0.608  & 0.642   & 0.672          & 0.666    & 0.661  & 0.639   & 0.648 & 52010.75         & 74.029    \\
8    & 0.633  & 0.701   & 0.684          & 0.640    & 0.649  & 0.636   & 0.657 & 344604.597       & 227.012   \\ 
\midrule
\rowcolor{lightgray} \multicolumn{10}{c}{Text-based Molecule Design (BLEU-2) - Tool (Name2SMILES,ChemDFM)} \\ 
\midrule
NUM  & Chain  & Random  & FullConnected  & Layered  & Star   & Debate  & Avg   & Avg\_all\_tokens & Avg\_Time \\ 
\midrule
0    & 0.832  & 0.832   & 0.832          & 0.832    & 0.832  & 0.832   & 0.832 & 2528.79          & 21.382    \\
1    & 0.842  & 0.842   & 0.842          & 0.842    & 0.842  & 0.842   & 0.842 & 3125.64          & 30.237    \\
2    & 0.844  & 0.844   & 0.844          & 0.844    & 0.844  & 0.844   & 0.844 & 18446.91         & 108.427   \\
4    & 0.864  & 0.888   & 0.885          & 0.867    & 0.884  & 0.889   & 0.880 & 150544.061       & 2207.641  \\
8    & 0.841  & 0.890   & 0.854          & 0.887    & \textbf{0.891}  & 0.874   & 0.873 & 733540.998       & 6744.788  \\ 
\midrule
\rowcolor{lightgray} \multicolumn{10}{c}{Text-based Molecule Design (BLEU-2) - Stacking Tool (Name2SMILES,ChemDFM)} \\ 
\midrule
NUM  & \multicolumn{6}{c}{Stacking Tool}                                                                 & Score & Avg\_all\_tokens & Avg\_Time \\ 
\midrule
2    & \multicolumn{6}{c}{{[}'ChemDFM\_2'{]}}                                                            & 0.898 & 2801.93          & 39.103    \\
3    & \multicolumn{6}{c}{{[}'Name2SMILES\_1','ChemDFM\_1'{]}}                                           & 0.918 & 2821.38          & 63.795    \\
4    & \multicolumn{6}{c}{{[}'Name2SMILES\_3','ChemDFM\_0'{]}}                                           & 0.904 & 2745.07          & 78.144    \\
4    & \multicolumn{6}{c}{{[}{[}'ChemDFM\_0', 'Name2SMILES\_1'{]}, 'ChemDFM\_1'{]}}                      & \textbf{0.925} & 2851.43          & 72.484    \\
8    & \multicolumn{6}{c}{\begin{tabular}[c]{@{}c@{}}{[}{[}'ChemDFM\_1','Name2SMILES\_1'{]},\\ {[}'ChemDFM\_1','Name2SMILES\_2'{]}{]}\end{tabular}} & 0.907 & 1830.11          & 95.623   \\
\bottomrule
\end{tabular}
    }
    \caption{Results of Multi-agent experiment. 0:FinalRefer agent is not included.}
    \label{tab:multiagent}
\end{table}
\label{appendix mulagent results}

\section{Case Study}
\label{casestudy}
As stated in Section \ref{6.4}, there are four distinct stacking behavior patterns-\textbf{correct}, \textbf{modify}, \textbf{judge} and \textbf{reserve} that make the stacking works. Here are three other cases for these patterns in the Tab.~\ref{tab:modify}:
\subsection{Correct}
As analyzed in Section \ref{6.4}, "Correct" is a very important pattern for agents. It was able to leverage the complementary strengths of multiple tools, significantly improving the accuracy of the final performance. 
\subsection{Modify}
As shown in the table, this is a common approach to using various tools. First, the problem is decomposed, and the RAG tool is used to retrieve information on each sub-question to obtain a preliminary answer. Then, subsequent processing is carried out using computational tools. Alternatively, one can first obtain an answer through computational tools, then have the agent self-assess the correctness of that answer, and finally use the RAG tool for cross-verify, thereby refining the answer and improving accuracy.
\setlength\tabcolsep{1pt}
\begin{table*}[!htb]
    \centering
    \tiny
    \definecolor{lightgray}{gray}{0.9} 
    \resizebox{1\textwidth}{!}{
    \begin{tabular}[l]{|>{\raggedright\arraybackslash}m{1.5cm}|>{\raggedright\arraybackslash}m{4cm}|>{\raggedright\arraybackslash}m{10cm}|}
    \hline
    \textbf{Stacking Pattern} & \textbf{Input} & \textbf{Output} \\ 
    \hline
    \textbf{Modify} 
    & The molecule is a peptide anion that is the conjugate base of gamma-Glu-Abu, obtained by removal of protons from the two carboxy groups as well as protonation of the amino group; major species at pH 7.3. It is a conjugate base of a gamma-Glu-Abu. Please try to infer the SMILES of this molecule.
    \newline
    \newline
    \newline
    \textcolor{red}{Gold Answer: CCC(NC(=O)CCC([NH3+])C(=O)[O-])C(=O)[O-]
}
    & To find the SMILES representation of the peptide anion that is the conjugate base of gamma-Glu-Abu, I first need to understand the structure of gamma-Glu-Abu. Gamma-Glu-Abu stands for gamma-glutamyl-aminobutyric acid. I will start by finding the SMILES of gamma-Glu-Abu.
    \newline
    \textbf{Action:} Name2SMILES  
    \newline
    \textbf{Input:} gamma-Glu-Abu
    \newline
    \textbf{Output:} CCC(NC(=O)CCC(N)C(=O)O)C(=O)O  \textcolor{red}{\ding{55}}
    \newline
    \textbf{Observation:} The next step is to modify this SMILES to represent the conjugate base of gamma-Glu-Abu at pH 7.3, where the two carboxyl groups are deprotonated and the amino group is protonated. This involves changing the carboxyl groups from "-C(=O)O" to "-C(=O)[O-]" and the amino group from "-N" to "-[NH3+]".
    \newline
    \textbf{Action:} ChemDFM  
    \newline
    \textbf{Input:} Modify the "CCC(NC(=O)CCC([NH3+])C(=O)[O-])C(=O)[O-]
"  from "-C(=O)O" to "-C(=O)[O-]" and the amino group from "-N" to "-[NH3+].
    \newline
    \textbf{Output:} CCC(NC(=O)CCC([NH3+])C(=O)[O-])C(=O)[O-] 
    \newline
    \textbf{Thought:} I now know the final answer
    \newline
    \textbf{Final Answer: CCC(NC(=O)CCC([NH3+])C(=O)[O-])C(=O)[O-]  \textcolor{red}{\ding{52}}}\\
    \hline
    \textbf{Judge}&The molecule is a cycloalkane composed of three carbon atoms to form a ring. It has a role as an inhalation anaesthetic. It is a cycloalkane and a member of cyclopropanes.Please try to infer the SMILES of this molecule.
    \newline
    \newline
    \newline
    \textcolor{red}{Gold Answer: CC(=O)N[C@@H]1[C@@H](O[C@@H]2O[C
    @@H](C)[C@@H](O)[C@@H](O)[C@@H]2
    O)[C@H](O[C@@H]2O[C@H](CO)[C@H](O
    )[C@H](O)[C@H]2O[C@@H]2O[C@@H](C)
    [C@@H](O)[C@@H](O)[C@@H]2O)[C@@
    H](CO)O[C@H]1O
}&The molecule described is a complex carbohydrate structure involving alpha-L-Fucp, beta-D-Galp, and D-GlcNAc units. The specific linkages and configurations are crucial for determining the correct structure. Given the complexity of the molecule, the best approach is to use the Molecule\_Design tools, which are designed to handle detailed questions about molecular structures. I will start with Molecule\_Design\_1 to see if it can provide a correct SMILES representation based on the detailed description provided.
\newline
\textbf{Action:} Molecule\_Design\_1
\newline
\textbf{Action Input:} The molecule is an alpha-L-Fucp-(1->2)-beta-D-Galp-(1->3)-[alpha-L-Fucp-(1->4)]-D-GlcNAc where the glucosamine at the reducing end has beta-configuration at its anomeric centre. It has a role as an epitope. Please provide the SMILES of this molecule.
\newline
\textbf{Observation:}CC(=O)N[C@@H]1[C@@H](O[C@@H]2O[C@@H](C)[C@@H](O)[C@@H](O)[C@@H]2O)[C@H]
(O[C@@H]2O[C@H](CO)[C@H](O)[C@H](O)[C@H]2O[C@@H]2O[C@@H](C)[C@@H](O)[C@@H](O)[C@@
H]2O)[C@@H](CO)O[C@H]1O
\newline
\textbf{Thought:}The SMILES provided by Molecule\_Design\_1 seems to represent a complex carbohydrate structure, but to ensure accuracy, I will use Molecule\_Design\_2 for a second opinion. This will help verify the correctness of the SMILES or provide an alternative if there are discrepancies.
\newline
\textbf{Action:} Molecule\_Design\_2
\newline
\textbf{Action Input:} The molecule is an alpha-L-Fucp-(1->2)-beta-D-Galp-(1->3)-[alpha-L-Fucp-(1->4)]-D-GlcNAc where the glucosamine at the reducing end has beta-configuration at its anomeric centre. It has a role as an epitope. Please provide the SMILES of this molecule.
\newline
\textbf{Observation:}Nc1ccn([C@@H]2O[C@H](COP(=O)([O-])[O-])[C@@H](O)[C@H]2O)c(=O)n1
\newline
\textbf{Thought:}The SMILES provided by Molecule\_Design\_2 is significantly different from the one provided by Molecule\_Design\_1, indicating a discrepancy. The structure from Molecule\_Design\_2 seems less likely to represent the complex carbohydrate structure described in the question. The SMILES from Molecule\_Design\_1 appears more plausible as it includes multiple sugar units linked together, which aligns better with the description of the molecule involving alpha-L-Fucp, beta-D-Galp, and D-GlcNAc. Therefore, I will consider the SMILES from Molecule\_Design\_1 as the more likely correct representation.
\newline
\textbf{Thought:} I now know the final answer
\newline
\textbf{Final Answer: CC(=O)N[C@@H]1[C@@H](O[C@@H]2O[C@@H](C)[C@@H](O)[C@@H](O)[C@@H]2O)
[C@H](O[C@@H]2O[C@H](CO)[C@H](O)[C@H](O)[C@H]2O[C@@H]2O[C@@H](C)[C@@H](O)[C@@H]
(O)[C@@H]2O)[C@@H](CO)O[C@H]1O\textcolor{red}{\ding{52}}}\\ \hline
\textbf{Reserve}&The molecule is a linear amino tetrasaccharide composed of N-acetyl-beta-D-galactosamine, alpha-D-galactose, beta-D-galactose and N-acetyl-beta-D-galactosamine units joined by sequential (1->3)-, (1->4)- and (1->4)-linkages. It has a role as an epitope.Please try to give SMILES of this molecule.
    \newline
    \newline
    \newline
    \textcolor{red}{Gold Answer: CC(=O)N[C@@H]1[C@@H](O)[C@H](O[C@
    @H]2O[C@H](CO)[C@H](O[C@H]3O[C@H]
    (CO)[C@H](O)[C@H](O[C@@H]4O[C@H](C
    O)[C@H](O)[C@H](O)[C@H]4NC(C)=O)[C@
    H]3O)[C@H](O)[C@H]2O)[C@@H](CO)O[C
    @H]1O
}&The SMILES representation for the molecule described is not readily available due to the complexity of the glycosidic linkages and the specific stereochemistry involved. Further detailed chemical synthesis knowledge or specialized software would be required to accurately construct the full SMILES notation of this tetrasaccharide.
\newline
\textbf{Thought:}The initial attempt using the Molecule\_Design\_2 tool did not yield a direct SMILES result, likely due to the complexity of the molecule described. As an alternative approach, I can attempt to construct the molecule step-by-step using the Name2SMILES tool for each individual sugar unit and then logically deduce the possible structure of the entire molecule. However, constructing the full SMILES for the entire molecule including specific linkages manually might not be feasible without specialized chemical software. I will first retrieve the SMILES for each sugar unit.
\newline
\textbf{Action:} Name2SMILES
\newline
\textbf{Action Input:} N-acetyl-beta-D-galactosamine
\newline
\textbf{Observation:} CC(=O)N[C@@H]1[C@@H](O)[C@@H](O)[C@@H](CO)O[C@H]1O
\newline
\textbf{Thought:}I have obtained the SMILES for N-acetyl-beta-D-galactosamine. Next, I will retrieve the SMILES for alpha-D-galactose.
\newline
\textbf{Action:} Name2SMILES
\newline
\textbf{Action Input:} alpha-D-galactose
\newline
\textbf{Observation:}OC[C@H]1O[C@H](O)[C@H](O)[C@@H](O)[C@H]1O
\newline
\textbf{Thought:}I have obtained the SMILES for alpha-D-galactose. Next, I will retrieve the SMILES for beta-D-galactose.
\newline
\textbf{Action:} ChemDFM
\newline
\textbf{Action Input:} What is the SMILES of beta-D-galactose
\newline
\textbf{Observation:} OC[C@H]1O[C@@H](O)[C@H](O)[C@@H](O)[C@H]1O
\newline
\textbf{Thought:}I have obtained the SMILES for beta-D-galactose. Since the molecule starts and ends with N-acetyl-beta-D-galactosamine, and I already have the SMILES for this unit, I do not need to retrieve it again. However, constructing the full SMILES string for the tetrasaccharide with specific glycosidic linkages (1->3, 1->4, 1->4) between these units is beyond the capability of the Name2SMILES tool and requires specialized chemical synthesis software or expertise in carbohydrate chemistry. Therefore, I will conclude my process.
\newline
\textbf{Thought: }I now know the final answer
\newline
\textbf{Final Answer: Unable to provide the exact SMILES string for the described tetrasaccharide due to the complexity of the glycosidic linkages and the limitations of the tools available.}\\ \hline
    \end{tabular}
    }
    \caption{Case analysis of some behavior patterns of agents during tools stacking.}
    \label{tab:modify}
\end{table*}

%
\subsection{Judge}
Judge refers to the process of selecting between two candidate answers based on the model’s knowledge in chemistry. This usually happens when two agent tools are available. When confronted with two anonymous tools that have the same descriptions, the model often opts to call both tools simultaneously before making a judgment. Furthermore, when using GPT-4o as the agent model, it typically demonstrates excellent judgment abilities.
\subsection{Reserve}
Reserve is a relatively uncommon phenomenon for large language models. Due to the hallucination problem, they usually choose to cover up knowledge blind spots to generate answers. However, when faced with multiple tools, the agent faces several answers. After thinking, it does not judge which one is more suitable, so it chooses to reserve opinions. Although this will reduce accuracy, in some ways it is a good thing to emerge this ability.

\end{document}